\documentclass[preprint,superscriptaddress,
amsmath,amssymb,aps,pre]{revtex4-2}
\usepackage{graphicx}
\usepackage{dcolumn}
\usepackage{bm}
\usepackage{color}
\usepackage{enumitem}
\usepackage{hyperref}
\hypersetup{linktoc=page}
\hypersetup{
    colorlinks,
    citecolor=blue,    filecolor=blue,
    linkcolor=blue,     urlcolor=blue
}
\graphicspath{{graphics/cavity/}{graphics/energy/}{graphics/fields/}{graphics/theory/}{graphics/distributions/}}

\begin{document}
\title{SRS-SBS competition and nonlinear laser energy absorption in a high temperature plasma}
\author{S. A. Shekhanov} 
\email{sviatoslav.shekhanov@eli-beams.eu}
\affiliation{ELI-Beamlines Center, Institute of Physics, Czech Academy of Sciences, 25241 Doln{\'i} B\v{r}e\v{z}any, Czech Republic}
\affiliation{Faculty of Nuclear Sciences and Physical Engineering, Czech Technical University in Prague, CZ-115 19 Prague, Czech Republic}
\author{V. T. Tikhonchuk}
\affiliation{ELI-Beamlines Center, Institute of Physics, Czech Academy of Sciences, 25241 Doln{\'i} B\v{r}e\v{z}any, Czech Republic}
\affiliation{Centre of Lasers  Intenses and Applications, University of Bordeaux, CNRS, CEA, 33405 Talence, France}
\date{\today}

\begin{abstract}
Stimulated Raman and Brillouin scattering of laser radiation in a plasma corona are outstanding issues for the inertial confinement fusion. Stimulated Raman scattering may produce absorption of a significant fraction of laser energy near the plasma quarter critical density associated with plasma cavitation and generation of hot electrons. By contrast, stimulated Brillouin scattering operates in a lower density plasma and prevents the laser light access to the absorption region. In the present paper, we report the results of analysis of competition of these two parametric instabilities with a series of one-dimensional kinetic simulations of laser-plasma interactions. By controlling the Brillouin backscattering through variation such plasma parameters as ion acoustic wave damping, divergence of the plasma expansion velocity or the laser bandwidth, we demonstrate the possibility of controlling the level of nonlinear laser absorption and scattering in a hot, weakly collisional plasma.
\end{abstract}

\maketitle

\section{Introduction}\label{sec:intro}

An important topic in inertial confinement fusion are the effects caused by various parametric instabilities generated by interaction of intense electromagnetic wave with hot dense plasma. It is important to control the interplay between scattering instabilities (stimulated Brillouin and Raman scattering) and laser absorption. Efficient absorption of the laser spike in the shock ignition scheme \cite{Betti_2007} is a serious and unresolved issue. Laser pulse propagates through a long and hot plasma corona created by the preceding compressing laser pulse, collisional absorption is very low in such plasmas and collective absorption related to excitation of parametric instabilities could be important. However, it is not known how much energy can be absorbed and reflected and how absorbed energy is distributed between thermal and suprathermal (hot) particles.

Studies of the laser-plasma interaction in the shock ignition conditions \cite{Batani_2014} show that stimulated Raman scattering (SRS) -- scattering of laser radiation on electron plasma waves -- is a dominant process resulting in producing a large quantity of hot electrons. These electrons can be absorbed in cold fuel and increase the shock strength \cite{Guskov_2012, Ribeyre_2013}, but there is a risk that a fraction of these electrons penetrates upstream the shock, preheats the plasma and reduces the shock strength. The energy distribution of hot electrons is thus needs to be controlled in the shock igniiton scenario.

Another issue is related to absorption efficiency of the laser spike. Stimulated Brillouin scattering (SBS) -- scattering of laser radiation on low frequency ion acoustic waves -- may prevent penetration of laser radiation into a dense plasma where SRS is excited, and thus compromise the efficiency of laser absorption. Large amounts of SBS were observed in experiments but do not fully explained theoretically \cite{Depierreux_2012}.

While multidimensional simulations are certainly closer to the reality, the simulated volume and simulation time are limited, the spatial and temporal resolution could be insufficient and parameter space cannot be studied in detail. For these reasons, one-dimensional (1D) simulations are still of great importance assuming that the dominant physical processes are correctly described. This applies to the physics of laser plasma interaction in the shock ignition conditions, where the main competitor of SRS -- two plasmon decay (TPD) \cite{Simon_1983} -- is suppressed because of high plasma temperature and interaction is dominated by SRS and SBS competition \cite{Gu_2019}.

Simplified theoretical models considering each parametric instability separately \cite{Liu_1974} are necessary for understanding the conditions where they can be excited but such models are insufficient for predicting the effects taking place in realistic conditions where several instabilities are interrelated and competing for same laser source. Moreover, secondary effects such as density profile modification and hot electron generation modify the conditions of instability excitation and subsequent nonlinear effects. It was demonstrated recently that two plasmon decay instability and stimulated Raman scattering could produce anomalous absorption of the incident laser light \cite{Turnbull_2020a, Maximov_2020}. Full kinetic simulations using particle-in-cell (PIC) codes are widely developed. Starting from simplified 1D geometry \cite{Estabrook_1976, Weber_2005, Klimo_2010}, contemporary PIC codes provide possibilities for extended 2D simulations \cite{Klimo_2013, Gu_2019} and 3D simulations are also accessible \cite{Xiao_2018, Wen_2019, Gu_2021}.

For these reasons we return in this paper to analysis of SBS-SRS competition in 1D geometry. Compared to previous studies \cite{Klimo_2010, Xiao_2018, Follett_2019, Wen_2021}, we focus our attention on the aspects that were observed but not investigated in detail. These are: (i) dynamics of SRS evolution leading to plasma cavitation and hot electron generation; (ii) development of SBS in a lower density plasma and (iii) controlling of SBS-SRS competition by changing the plasma expansion velocity, ion species composition and the laser pulse bandwidth. The cavitation process has been reported in the 1970s \cite{Estabrook_1976}, but it role was underestimated. Other publications has reported on cavitation due to SBS \cite{Weber_2005} and SRS \cite{Klimo_2010, Moreau_2017}, but its relation to laser energy absorption and hot electron generation was not quantified. These studies show that cavitation is a promising process for increasing of a nonlinear laser energy transfer to electrons with energies in the range of $10-20$ keV, which are favorable for shock ignition. 

The paper is organized as follows. The theoretical background is recalled in section \ref{sec:theory}. Numerical model and results are presented in section \ref{sec:model}. We are using the PIC code SMILEI \cite{Smilei_2018} in the collisionless version. However, several results, which take into account electron collisions, are also presented. They are important for evaluation of the SRS process near the quarter critical density where the scattered wave is excited near its turning point. The results are discussed and concluded in section \ref{conclusion}.

\section{Theoretical analysis}\label{sec:theory}
\subsection{Temporal and spatial characteristics of SBS in a linear regime}\label{sec:sbs}
Stimulated Brillouin scattering is a three-wave parametric instability corresponding to a decomposition of the pump electromagnetic wave with a frequency $\omega_0$ into a scattered electromagnetic wave and an ion acoustic wave with a frequency $\omega_{ia} \ll \omega_0$. In the case of backward scattering, the characteristic temporal growth rate of SBS reads \cite{Liu_1974, Kruer_1988}:
\begin{equation}\label{eq:gamma}
\gamma_{\rm SBS} = \frac{1}{4}\omega_{pi} v_{os} \sqrt{\frac{k_{ia}}{\omega_0 c_s}},
\end{equation} 
where $\mathbf{k}_{ia}=\mathbf{k}_{0}-\mathbf{k}_{s} \approx 2\mathbf{k}_{0}$ is the ion acoustic wave vector, $\mathbf{k}_{0}$ and $\mathbf{k}_{s} \approx -\mathbf{k}_{0}$ are the wave vectors of the pump and scattered waves, $k_0=(\omega_0/c)\,(1-n_e/n_c)^{1/2}$, $\omega_{pi}$ is the ion plasma frequency, $\omega_{ia}=k_{ia} c_s$ is ion acoustic wave frequency, $c_s$ is the ion acoustic velocity, $c$ is the light speed, $n_e$ is the local electron density, $n_c$ is the critical density for the laser light and $v_{os}$ is the amplitude of the electron quiver velocity in the laser field. 

The temporal gain is reduced if the pump wave has the finite spectral bandwidth $\Delta\omega_0\gtrsim \gamma_{\rm SBS}$ \cite{Thomson_1974, Thomson_1975, Laval_1977, Pesme_2007}. However even in the case $\Delta\omega_0 \lesssim \gamma_{\rm SBS}$ one can expect a decrease of temporal gain. The growth rate can be estimated with approximate expression:
\begin{equation}\label{eqDeltaomega}
 \gamma_{\Delta\omega_0}\simeq \frac{\gamma_{\rm SBS}^2}{\sqrt{\gamma_{\rm SBS}^2 + \xi^2 \Delta\omega_0^2}},
\end{equation}
where the numerical coefficient $\xi$ in the right hand side depends on the pump wave power spectrum.
The temporal gain is also affected by the ion acoustic wave damping $\gamma_{ia}$, which is of particular importance in a multi-ion species plasma \cite{Bychenkov_1994, Bychenkov_1995a, Williams_1995, Kozlov_2002, Feng_2019}.

SBS is a convective instability, the amplitude of scattered wave in an inhomogeneous plasma is amplified spatially along its propagation direction near the resonant point corresponding to the perfect wave matching. The spatial growth rate for SBS reads \cite{Liu_1974, Kruer_1988, Hinkel_2008}:
\begin{equation}\label{gainSRS}
G_{\rm SBS} = \frac{1}{4}\frac{k_{ia}^2\,v_{os}^2}{v_{s}\,\omega_s}\int dx\,{\rm Im} \frac{\chi_{e}\,(1 + \chi_{i})}{1+\chi_e + \chi_i},
\end{equation}
where the integral is taken over the resonance region, $v_{s}$ is the group velocity of scattered wave and $\chi_{e,i}$ is the electron/ion susceptibility corresponding to the ion acoustic wave.

\begin{figure}[!ht]
\includegraphics[width=0.45\linewidth]{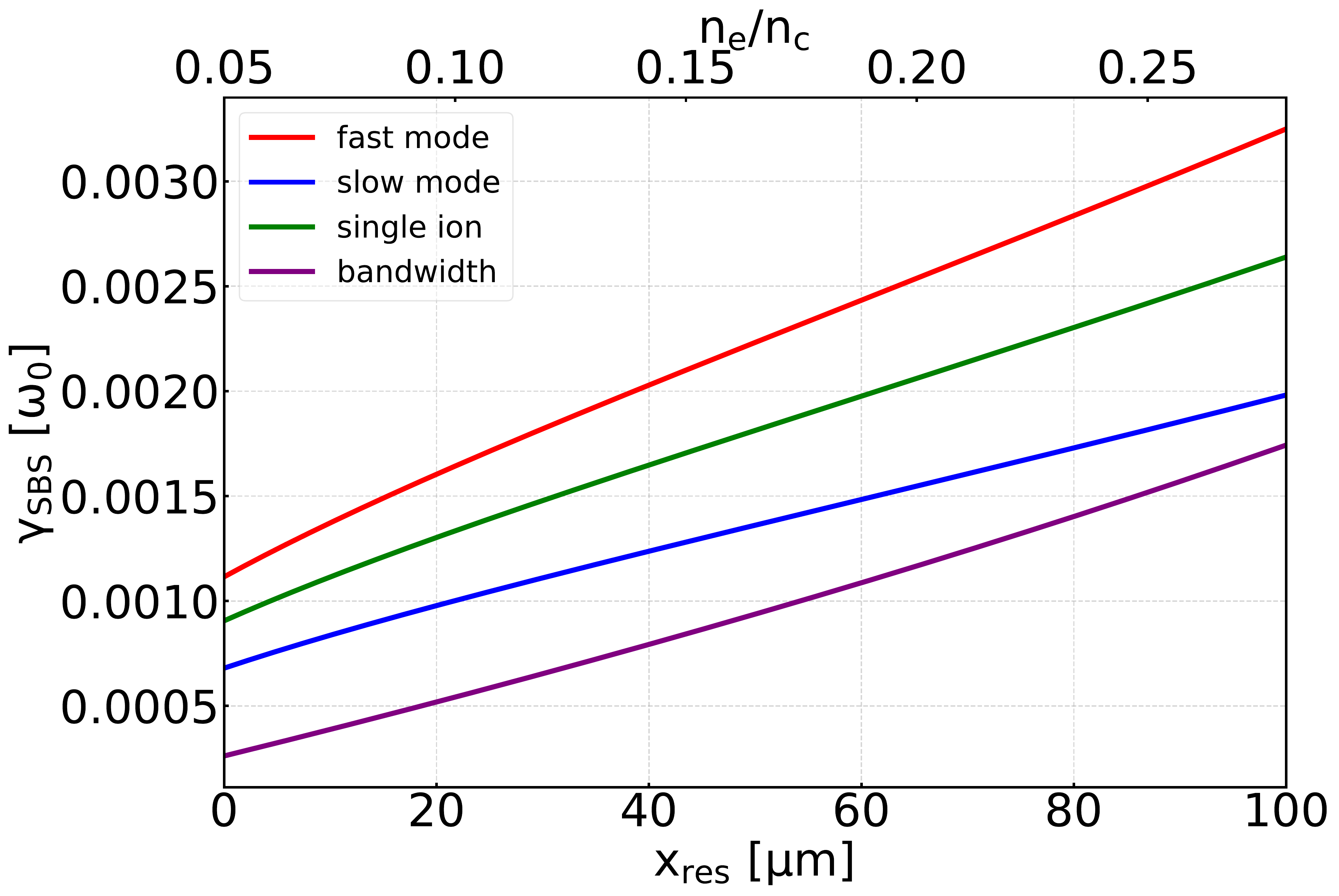}\hspace{3mm}
\includegraphics[width=0.43\linewidth]{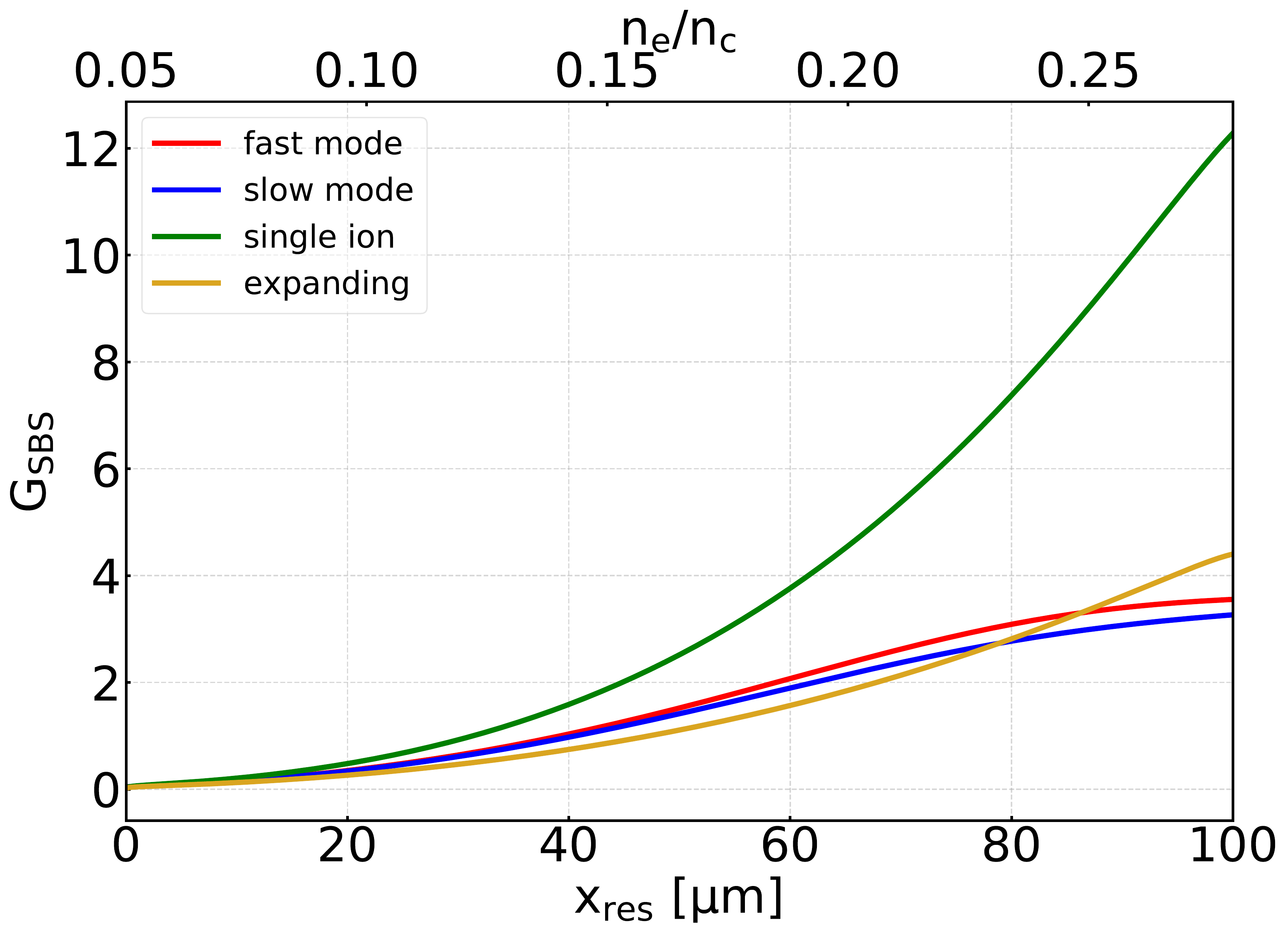}
\caption{Dependence of the growth rate (a) of SBS on the resonance position and spatial gain (b) on the coordinate for the case of a plasma with inhomogeneous density profile with the density scale length $L_n=300\,\lambda_0$ and velocity scale length $L_u=300\,\lambda_0$. Here, the laser intensity is $6\times10^{15}$~W/cm$^2$, laser wavelength is $0.351\,\mu$m, electron temperature is 3.4 keV and ion temperature is 1~keV.} \label{fig:gains}
\end{figure}

Examples of the SBS temporal growth rate dependence on the plasma density for the set of parameters considered in numerical simulations presented in the next section are shown in figure \ref{fig:gains}a for a plasma with the electron density increasing linearly with the coordinate $x$ according to the relation $n_e/n_c = 0.05 + 0.23\,x/L_n$. Here, $n_c=1.1\times 10^{21}\lambda_0^{-2}$ cm$^{-3}$ is the electron critical density for the pump wave, $\lambda_0 =0.351\,\mu$m is the pump wavelength and $L_n =300\,\lambda_0$ is the characteristic density scale length. The red, green and blue lines show the SBS growth rate in a plasma with fully ionized carbon and hydrogen ions of equal concentration, $n_C=n_H=\frac17n_e$. The green line shows the case where two ion species are replaced by one species with an average mass $A_{av}=m_{i,av}/m_p=6.5$ and charge  $Z_{av}=3.5$, where $m_p$ is the proton mass. There is a significant difference from the case where two ion species are considered separately. There are two ion acoustic modes: the slow mode is strongly damped due to the resonant interaction with ions and, consequently, its growth rate is suppressed (blue line). By contrast, the fast mode is less damped and it has a significantly larger growth rate (red line).

The purple line represents the growth rate for a phase-modulated laser pulse with a Lorentzian power spectrum with a correlation time $\tau_c = 2/\Delta\omega_0\sim 0.5$ ps in the case of a single average ion species. The laser bandwidth $\Delta\omega_0 \simeq 4$ ps$^{-1}$ is comparable with the growth rate  $\gamma_{sbs} \sim 5-10$ ps$^{-1}$ in the monochromatic pump shown in figure \ref{fig:gains}a (green line) and, according to equation \eqref{eqDeltaomega}, the growth rate has suppressed.

The golden line in figure \ref{fig:gains}b represents the spatial gain for an inhomogeneous plasma with the ion flow velocity $u$ depending on coordinate. Here we consider a single ion model and a linear flow velocity profile, $u(x) = u_s\,(x/L_u - 1)$ where $L_u=300\,\lambda_0$ is the velocity scale length and $u_s = 0.75\,\mu$m/ps is approximately twice the ion sound velocity $c_s \simeq 0.47\,\mu$m/ps. In this case the frequency of ion acoustic wave is Doppler-shifted, $\omega_{ia} = k_{ia}(c_s + u)$, and the resonance condition, $\kappa(x) = k_0 - k_s - k_{ia}\approx 0$, is fulfilled in a narrow zone depending on the density, $L_n=n_e\,(dx/dn_e)$, and velocity, $L_u=u\,(dx/du)$, scale lengths. Consequently, amplification of SBS-driven waves is suppressed if the plasma expansion is considered. This can be seen by comparing the golden line in figure \ref{fig:gains}b with the green line corresponding to zero flow velocity. 

Ion acoustic wave damping has a smaller effect on the spatial SBS gain as shown in figure \ref{fig:gains}b. There is practically no difference between the spatial gains for slow and fast modes, but the gain for a plasma with a single average ion species is higher.  These examples show the possibilities of controlling the SBS by laser bandwidth and ion species ratio.

\subsection{Temporal and spatial characteristics of SRS in a linear regime}
\begin{figure}[!ht]
\includegraphics[width=0.45\linewidth]{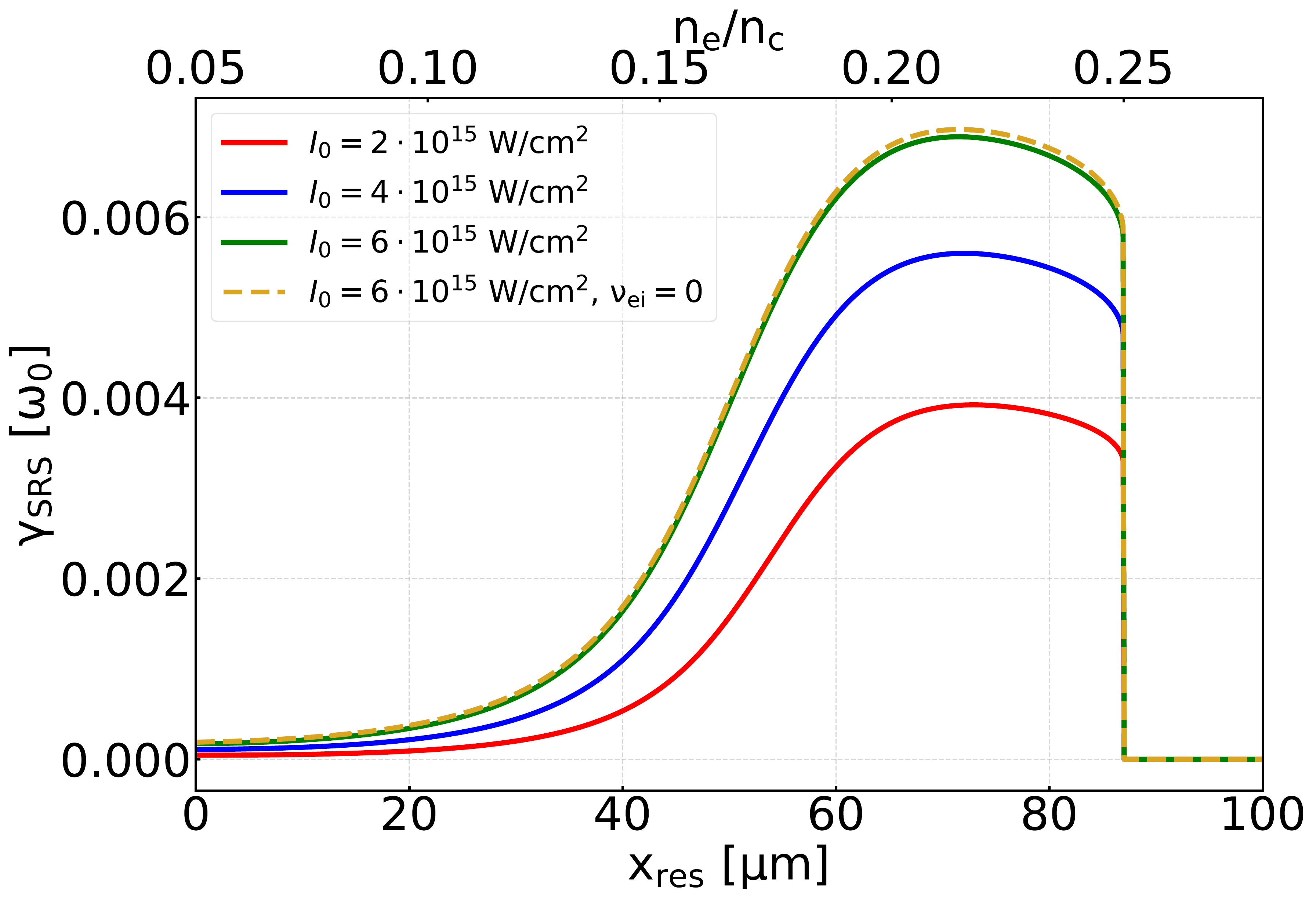}\hspace{3mm}
\includegraphics[width=0.47\linewidth]{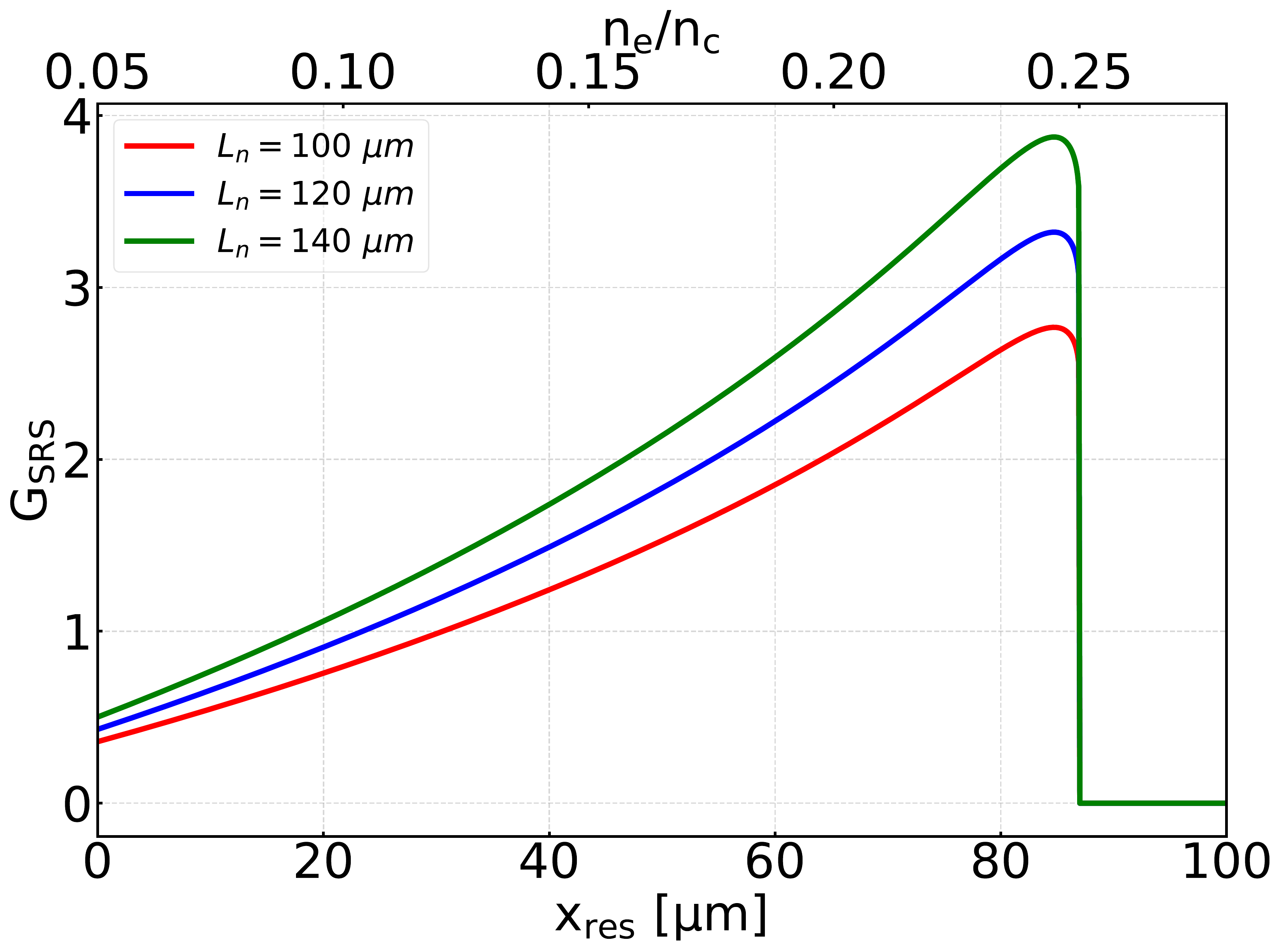}
\caption{Dependence of the SRS temporal growth rate (a) and spatial gain (b) on the coordinate of the resonance point for the case of a plasma with inhomogeneous density profile. The density scale length $L_n=300\,\lambda_0$ in case (a) and varies from 300 to $450\,\lambda_0$ in case (b). The laser intensity is $(2-6)\times10^{15}$~W/cm$^2$ in case (a) and $6\times10^{15}$ W/cm$^2$ in case (b). Here the laser wavelength is $0.351\,\mu$m, electron temperature is 3.4 keV and ion temperature is 1~keV.} \label{fig:gains_SRS}
\end{figure}

Stimulated Raman scattering is the parametric instability corresponding to a decomposition of the pump wave into a scattered electromagnetic wave and an electron plasma (Langmuir) wave. As the frequencies of plasma and scattered waves are comparable, SRS can be excited only in a plasma with electron density smaller than the quarter of critical density. This is an electronic instability characterized by a high growth rate \cite{Liu_1974, Kruer_1988} as shown in figure \ref{fig:gains_SRS}a
\begin{equation}\label{eq:gamma SRS}
\gamma_{\rm SRS} =\frac{k_p v_{os}}4 \sqrt{\frac{\omega_{pe}}{\omega_0-\omega_{pe}}},
\end{equation}
where $\mathbf{k}_{p}=\mathbf{k}_{0}-\mathbf{k}_{s}$ is the plasma wave vector. Similarly to SBS, SRS in the backward direction is a convective instability with the spatial gain
\begin{equation}\label{eq:gainSRS}
 G_{\rm SRS}= \frac{\pi}{8}\frac{v_{os}^2}{c^2}\frac{k_p^2L_n}{k_s}.
\end{equation}
which is shown in figure \ref{fig:gains_SRS}b. There is also an absolute SRS instability \cite{Liu_1974, Afeyan_1985, Follett_2019}, which develops near the quarter critical density corresponding to the turning point of the scattered wave. This absolute instability has a rather low threshold 
\begin{equation}\label{eq:absSRS}
v_{os}/c \simeq(k_0L_n)^{-2/3},
\end{equation}  
and it often dominates the interaction. The phase-modulation of the pump pulse with correlation time $\tau_c \sim 0.5$ ps does not affect on the SRS instability development because the bandwidth $\Delta\omega_0$ is much smaller than the SRS growth rate.

The SRS growth rate shown in figure \ref{fig:gains_SRS}a attends its maximum in a $\sim 10\,\mu$m wide region slightly below the quarter critical plasma density. In the lower density part of plasma, SRS is suppressed due to strong Landau damping because the characteristic parameter $k_p\lambda_D$ increases, were $\lambda_D$ is the Debye length. In our simulations, we have fixed the intensity at $6 \times 10^{15}$ W/cm$^2$ and density scale length at $L_n=300\,\lambda_0$. 

The SRS instability could significantly contribute the laser absorption via supra-thermal electrons with energies about ten times higher than the thermal energy. They could lead to fuel preheat in the inertial confinement fusion, so their energy distribution has to be controlled.

Collisional damping may be an important issue for the SRS-driven plasma waves. The scattered light group velocity, $v_g\simeq c\,(1-\omega_{pe}^2/\omega_s^2)^{1/2}$, could be small near the corresponding critical surface, and electron-ion collisions may contribute to the plasma wave damping \cite{Klimo_2010, Xiang_2011}. The minimal scattered wave group velocity near the turning point is defined by the limit of applicability of the WKB approximation, that is, $k_s \gtrsim (L_n\,\lambda_{D}^2)^{-1/3}$. The WKB theory of SRS was revisited near these turning points in Refs. \cite{Liu_1974, Afeyan_1985}. 

Collisional damping rate of the scattered electromagnetic wave is proportional to the electron-ion collision frequency evaluated at the quarter critical density $\nu_{ei}^*$, which is $\sim 2.1$ ps$^{-1}$ under the considered conditions. This value is smaller than the SRS temporal growth rate but could be comparable or larger than the collisionless damping rate of the electromagnetic wave trapped in the cavity  \eqref{cavity_absorption}, which is considered in the next section. Absorption of the scattered wave propagating from the quarter critical zone $x_{1/4}$ to the plasma edge reads:
\begin{equation}\label{ei_absorption}
f_{\rm coll.abs} = 1 - \exp\Bigl[-\frac{\nu_{ei}^*}{c}\int_0^{x_{1/4}}\,\Bigl(\frac{\omega_{pe}}{\omega_{s}}\Bigr)^4\,\Bigl(1 - \frac{\omega_{pe}^2}{\omega_{s}^2}\Bigr)^{-1/2}\, dx \Bigr].
\end{equation}
For the chosen plasma parameters the value of absorption coefficient is $f_{\rm col.abs}\approx 0.3$. Thus about 30\% of the scattered wave energy could be absorbed in plasma due to collisions.

\subsection{Collisionless SRS absorption in a cavity}
Numerical simulations described in the next section show that the absolute SRS instability may lead to a significant absorption of laser energy in plasma. It is related to formation of density cavities near the quarter critical density where a part of energy of scattered wave is trapped and absorbed by the electrons. As cavity expands under the action of pondermotive pressure of trapped wave, a part of energy is also transferred to ions. Here, we estimate the efficiency of absorption of electromagnetic waves trapped in a cavity.

\begin{figure}[!ht]
\includegraphics[width=0.45\linewidth]{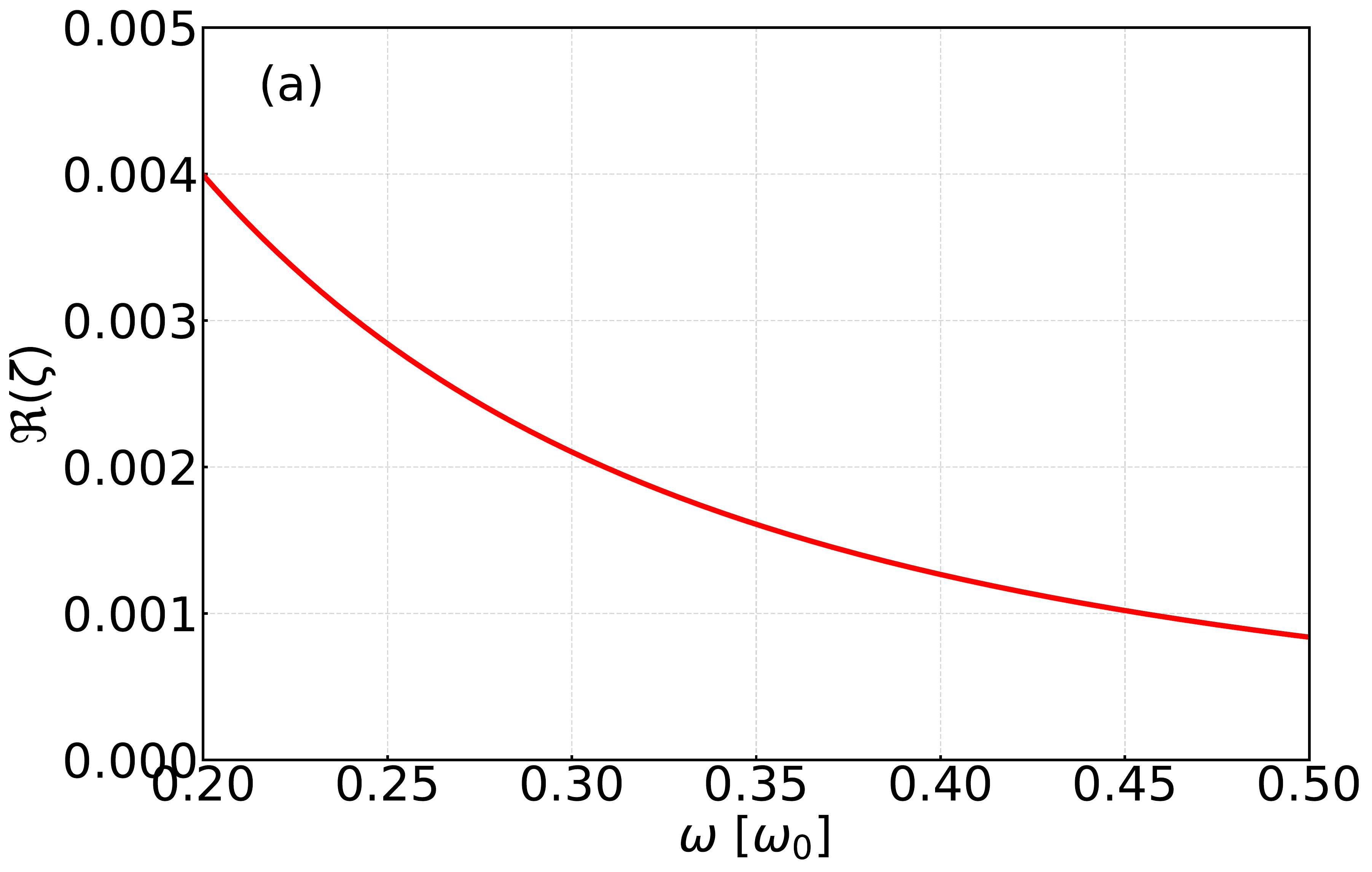} \hspace{3mm}
\includegraphics[width=0.45\linewidth]{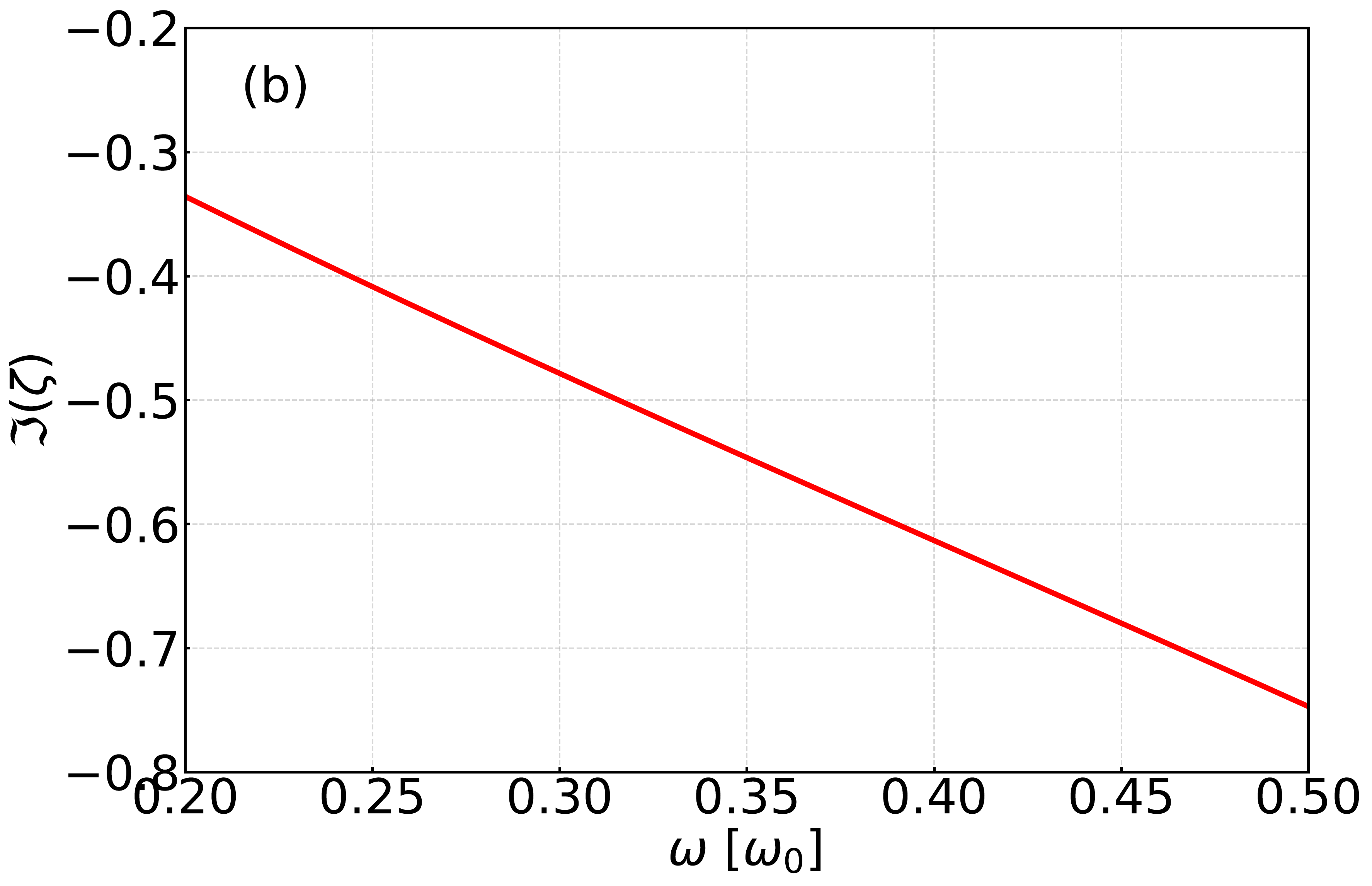}
\caption{Real (a) and imaginary (b) parts of the surface impedance for a representative set of parameters: $\Delta x_w/\lambda_0= 7.7$, $v_{Te}/c=0.082$, $\omega_{p, t}/\omega_0=0.49$ and $\omega_{b, t}/\omega_0= 0.255$.
}\label{fig:impedance}
\end{figure}

After a fast transient process, the cavity can be considered as a deep rectangular well of a width $\Delta x_w$, top density $n_{e,t}$  and bottom density $n_{e,b}$ filled with a resonant electromagnetic wave. The frequency of the trapped wave $\omega_w$ is defined by the dispersion equation:
\begin{equation}\label{eq7}
i\,\zeta(\omega_w) =R(\omega_w)= \sqrt{1 - \omega_{p,b}^2/\omega_w^2}\cot\left[(\Delta x_w/2c)\,\sqrt{\omega_w^2 - \omega_{p,b}^2}\right],
\end{equation}
which is obtained by imposing the reflecting boundary conditions at the cavity edges. Here, $\omega_{p,b}$ is the plasma frequency corresponding to the bottom well density $n_{e,b}$, $\zeta(\omega)$ is the surface impedance expressed through the Fried-Conte $Z$-function \cite{Fried_1961}:
\begin{equation}\label{impedance}
\zeta(\omega) = -\frac{2\sqrt{2}}{\pi}\,\frac{v_{Te}}{c}\int_{0}^{\infty}\frac{d\xi}{1-2\,\xi^2\,(v_{Te}/c)^2 - 2\,\xi^3\,(v_{Te}/c)^2\,(\omega_{p, t}/\omega)^2\,Z(\xi)},
\end{equation}
where $\omega_{p, t}$ is the plasma frequency corresponding to the top well density $n_{e, t}$ and $v_{Te}$ is electron thermal velocity. The real and imaginary parts of an impedance calculated numerically are shown in figure \ref{fig:impedance}.

Equation \eqref{eq7} has multiple solutions, the lowest frequency modes $\omega_w\gtrsim \omega_{p, t}$ are of main interest. For a representative set of parameters, $\Delta x_w/\lambda_0= 7.7$, $v_{Te}/c=0.082$, $\omega_{p, t}/\omega_0=0.49$ and $\omega_{b, t}/\omega_0= 0.255$, the real and imaginary parts of impedance are shown in figure \ref{fig:impedance}. The top plasma frequency corresponds to the quarter critical plasma density region where SRS occurs.  As the real part of impedance is much smaller than the imaginary part, the real part of the lowest proper frequencies $\omega_{w1}/\omega_0=0.258$ and  $\omega_{w2}/\omega_0=0.266$ can be found from equation \eqref{eq7} with only imaginary part of $\zeta$ retained, $-{\rm Im}\,\zeta(\omega_w)=R(\omega_w)$. 

Then, by taking the imaginary part of equation \eqref{eq7} one can find expression for the imaginary part of the proper frequency, ${\rm Im}\,\omega_w={\rm Re}\,\zeta(\omega_w)/R^\prime(\omega_w)$. The values of damping for both modes  ${\rm Im}\,\omega_{w}/\omega_0\simeq -1\times 10^{-4}$ correspond to the collisionless absorption of trapped modes due to the interaction with plasma electrons in the skin layers from both sides of the well. The absorption coefficient averaged by a laser period is defined as:
\begin{equation}\label{cavity_absorption}
f_{\rm cav.abs} =- 4\pi\,{\rm Im}\, \omega_w/{\rm Re}\, \omega_w.
\end{equation}
The value of $f_{\rm cav.abs}$ is of the order of 0.1\% for the chosen parameters. 

SRS provides a flexible way for controlling the nonlinear laser absorption. The collisional and collisionless processes compete each other since the damping rate of the trapped electromagnetic wave is compatible with the collision absorption. We explore this feature numerically in the next section \ref{sec:model}.

\section{Simulation model and results}\label{sec:model}
\subsection{Input parameters}\label{subsec:input}

In order to demonstrate the possibilities of controlling nonlinear laser energy absorption, we consider the interaction of an intense laser pulse incident normally on an expanding underdense inhomogeneous plasma made of a plastic (CH) with parameters relevant to the shock ignition scenario \cite{Betti_2007, Batani_2014}. Plasma has a linear density profile, $n_e/n_c=0.05+0.23\, x/L_n$, increasing from 0.05 to $0.28\,n_c$ over a length $L_n= 286\,\lambda_0$ for the laser wavelength  $\lambda_0=0.351\,\mu$m. The initial electron temperature, $T_e = 3.4$ keV, and ion temperature, $T_i = 1.0$ keV, are assumed to be homogeneous in space. These are representative values reported in numerical simulations and experiments \cite{Batani_2014, Gu_2019}. The laser pulse intensity, $I_{0}= \frac12 c\epsilon_0 E_{0}^2=6\times 10^{15}$ W/cm$^{2}$, is maintained constant during the simulation time of 8~ps, that is, about $10^4$ laser periods, after a linear ramp during first ten laser periods. This time is sufficient for achieving a quasi-steady state in the simulation.

Simulations were performed with a massively parallel fully-relativistic electromagnetic PIC code SMILEI \cite{Smilei_2018} in a 1D geometry. The code uses smooth high-order shape functions for particles in order to suppress numerical heating and reduce the number of particles per cell. The number of particles per cell in our simulation is 1500, which is sufficient for maintaining a low noise level in a 1D simulation. The simulation cell size is $dx =0.02\,\lambda_0$ and the computational time step is $dt= 0.02\,\lambda_0/c$. 

The simulation box length is about $314\,\lambda_0$ with $14\,\lambda_0$ vacuum margins at the front and rear sides of the box in order to enable a free plasma expansion. Boundary conditions are open for electromagnetic waves, the particles reaching front and rear boundaries are re-injected with a Maxwellian distribution corresponding to the initial temperature. 
These boundary conditions, however, do not describe ejection of hot electrons into a dense plasma as it happens in the experiment. Instead, we evaluate the distribution of the energy flux carried by electrons and ions in plasma near the right boundary.
 
In order to evaluate the SBS-SRS competition and its role in the efficiency of laser energy absorption, we consider six representative cases:
\begin{enumerate}[label=\roman*)]\itemsep-5pt
\item the reference case of the interaction of a monochromatic laser pulse with a collisionless plasma with zero fluid velocity and a single ion species having effective charge $Z_{av}=3.5$ and mass $6.5\,m_p$ corresponding to the equimolar mixture of hydrogen and carbon; 
\item the reference case is repeated with the electron-ion collisions switched on;
\item the same as case i) but with expanding plasma with a fluid velocity linearly increasing with the coordinate, $u(x)=u_s(x/L_u-1)$, with $L_u=300\lambda_0$ and $u_s=.075\,\mu$m/ps; 
\item the case iii) of expanding plasma is repeated with the electron-ion collisions switched on. 
\item the same as case i) but with two ion species, fully ionized carbon and hydrogen of equal concentrations; 
\item the same as the reference case i) but with a phase-modulated laser pulse with a correlation time $\tau_c\sim 0.5$ ps.
\end{enumerate}
The results are compared in terms of laser energy absorption in plasma, reflection and transmission, absorbed laser energy partition between electrons and ions, and energy partition between the bulk (thermal) and hot (suprathermal) particles.

The phase-modulated laser case was implemented using a random phase jump technique. The time of jump is fixed on 0.5 ps and the magnitude is random within the $2\pi$ interval. In this case the phase undergoes a random walk following the Poisson probability function. According to the theoretical analysis presented in section \ref{sec:theory}, the expected linear growth rate for SBS instability  shown in figure \ref{fig:gains}a  is $\gamma_{\Delta\omega_0} \sim 5.4$~ps$^{-1}$ for the laser bandwidth $\Delta\omega_0 \simeq 4$~ps$^{-1}$, which corresponds to the correlation time $\tau_c \sim 0.5$ ps. Thus, the laser pulse bandwidth is comparable with the SBS growth rate and may affect its development.

\subsection{Particle energy evolution}\label{subsec:energy}
In absence of electron-ion collisions, which are of minor importance under the conditions considered here, the energy transfer between the laser and the particles is mediated by electrostatic fields. In the case of SRS, the laser energy is transferred to electron plasma wave, which then transfers it to electrons, in addition, in the case of cavitation, the scattered electromagnetic wave is trapped and transfers a part of its energy to ions. The energy transfer in the case of SBS is weaker because of smallness of the ion acoustic wave frequency, but then, ion acoustic wave transfers its energy directly to ions.

\begin{figure}[!ht]
\includegraphics[width=0.45\linewidth]{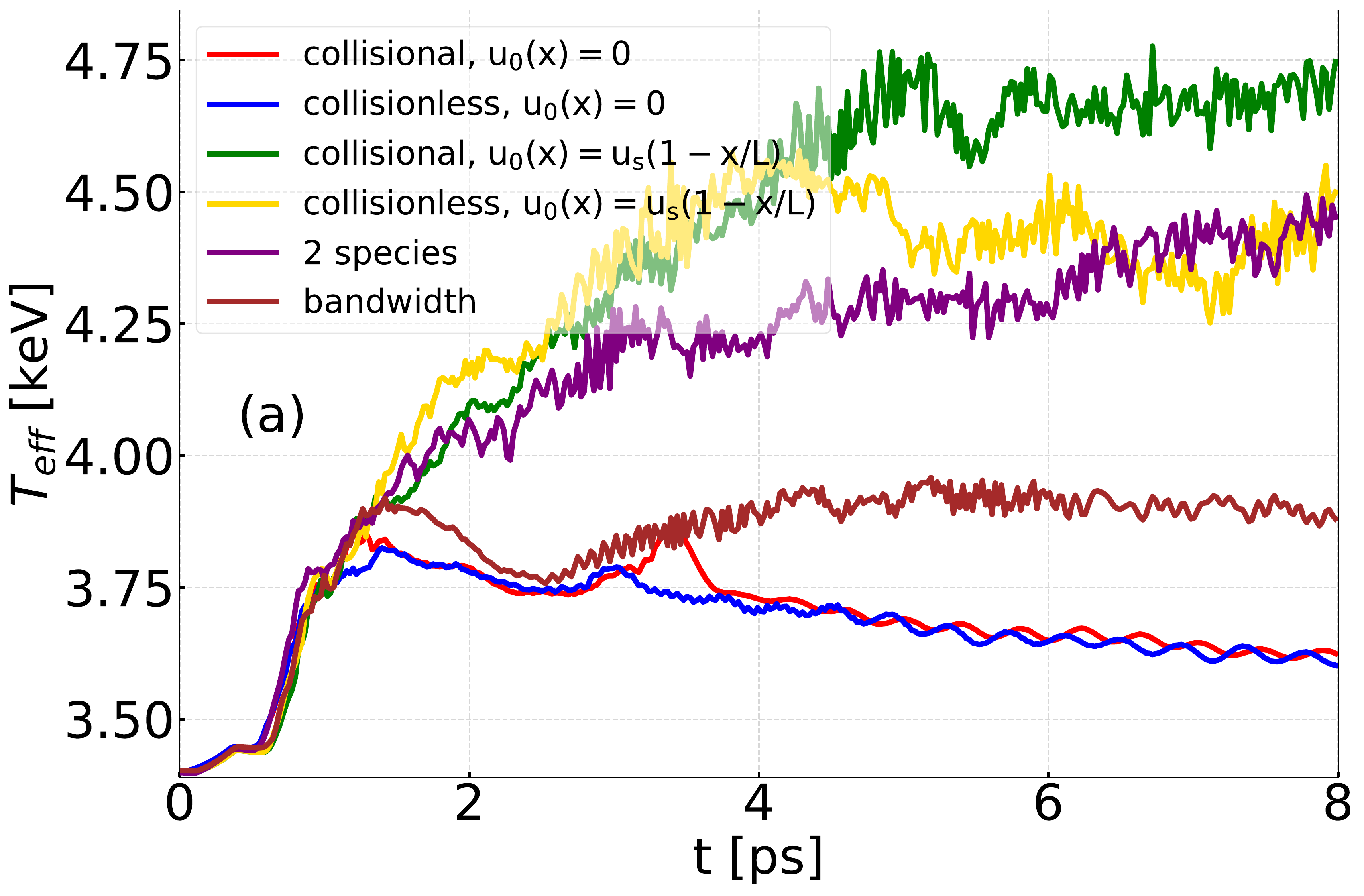}\hspace{3mm}
\includegraphics[width=0.45\linewidth]{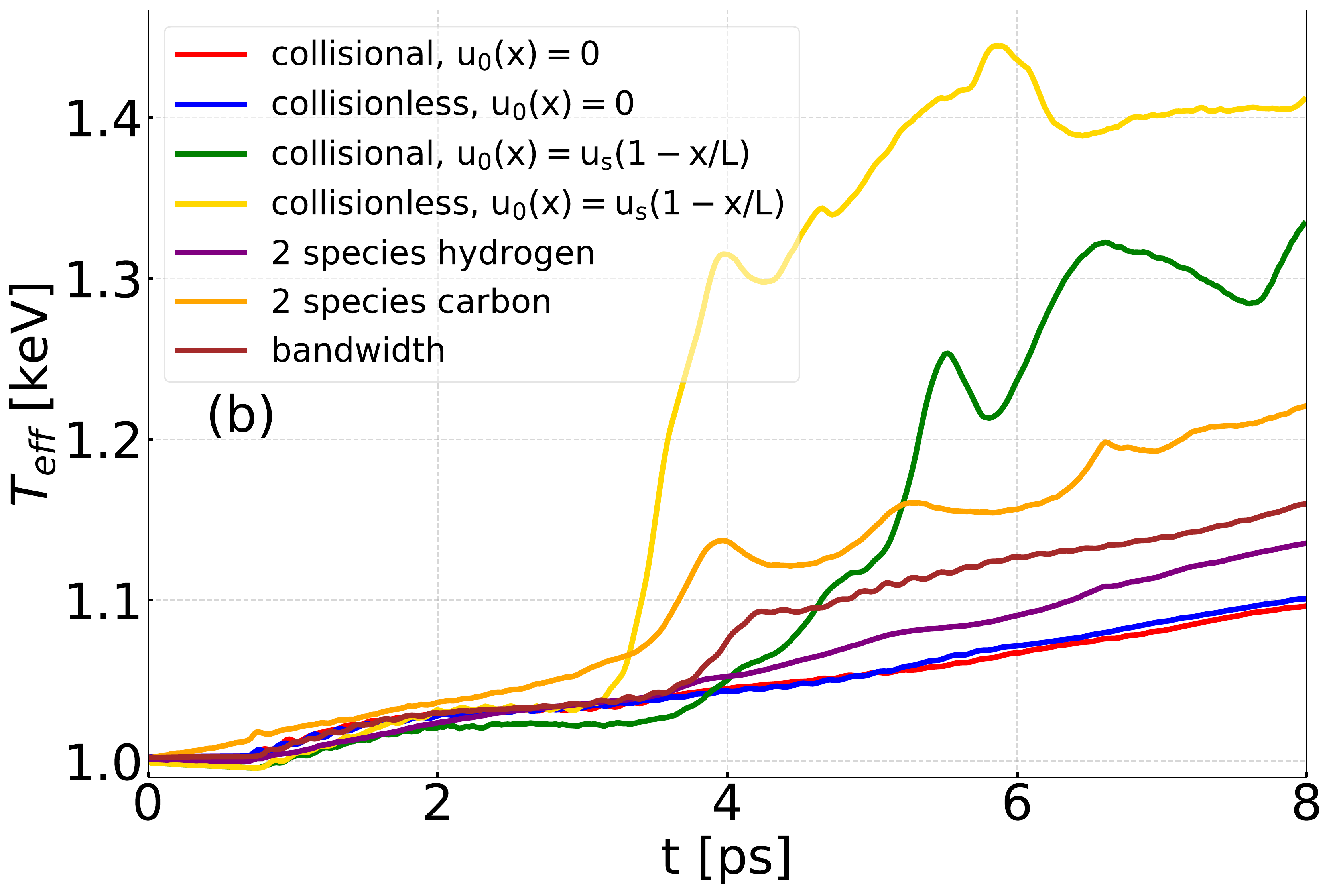}
\caption{Temporal evolution of the space averaged temperature of electrons (a) and ions (b) in the runs described in the legend. Parameters are given in section \ref{subsec:input}.}\label{fig_energy}
\end{figure}

The temperature of species $j=e,i$ at the position $x$ is defined via the distribution function $f_j$:
\begin{equation}\label{eq1}
T_{j} =\frac{2}{3n_j} \int d\mathbf{p}\, \varepsilon_j f_j(\mathbf{p}, x, t),
\end{equation}
where $\varepsilon_j=m_j c^2(\gamma_j-1)$ is the particle energy,  $\gamma_j$ is the relativistic factor, $m_j$ is the particle mass and $n_j=\int d\mathbf{p}\,f_j$ is the density.  Temporal evolution of the effective temperature of electrons and ions in the simulations averaged over the length of simulation box is shown in figure \ref{fig_energy}.

The particle energy starts evolving at $t = 0.7$ ps when laser pulse enters the plasma. The first fast increase of electron temperature, same in all simulations in figure \ref{fig_energy}a, is related to the excitation of SRS near the quarter critical density. After that, two different evolution paths are observed. In the reference case of a single species  plasma with zero velocity, the electron temperature gradually decreases independently on the presence (red) or absence (blue) of electron collisions. A comparison of the collisional and collisionless runs confirms the minor role played by collisions in the reference case. By contrast, the ion temperature shown in figure \ref{fig_energy}b increases at a later time of $\sim 3.5$ ps when the SBS instability is excited.

In the cases i) and ii) of a plasma with zero velocity shown in figure \ref{fig_energy} with red (collisional case) and blue (collisionless case), SBS dominates the interaction. After a short delay of $\sim 3$~ps a strong backscattering prevents the laser radiation to penetrate to the quarter critical density where SRS can be excited. Consequently, SRS is suppressed and electron temperature decreases. A small peak in the electron temperature near $t=3.5$ ps is explained by the collisional absorption of SRS reflected light excited near the quarter critical density.  As shown in table \ref{tab:table1}, only about 2\% of laser energy is absorbed in the reference case i) and distributed between electrons and ions. The level of SBS reflectivity is very high, about 93\% at the late stage of interaction ($t>5$ ps). 

\begin{table}[h!]
  \begin{center}
   \begin{tabular}{|c|c|c|c|c|} \hline
   \#  &  Simulation  & Reflected & Transmitted & Absorbed \\  \hline
    i) &  Reference, collisionless plasma  & 91\% & 4\% & 5\% \\ \hline
   ii) &  Reference, collisional plasma & 88\% & 10\% & 2\%  \\  \hline 
  iii) &  Expanding collisionless plasma & 36\% & 37\% & 27\% \\  \hline
  iv)  &  Expanding collisional plasma & 37\% & 27\% & 36\% \\  \hline
   v)  &  Two ion species, collisionless  & 36\% & 30\%      & 34\%   \\ \hline
  vi)  &  Phase-modulated pump, collisionless  & 81\% & 16\% & 3\%  \\  \hline    \end{tabular}
 \caption{Averaged values of reflected and transmitted light for the last 4 ps of simulation time for the six considered cases.}\label{tab:table1}
  \end{center}
\end{table}

In the cases iii) and iv) of expanding plasma shown in figure \ref{fig_energy} with green (collisional case) and gold (collisionless case) lines, electron and ion temperatures are increasing to higher levels, thus demonstrating a much better laser absorption. As discussed in section \ref{sec:theory}, divergence of the plasma velocity suppresses SBS as the amplification length of the scattered wave is shorter. Consequently, laser absorption is dominated by SRS near the quarter of critical density and plasma cavitation.  Similar effect has been already observed by Klimo et al. \cite{Klimo_2010}. In our case about 9\% of laser energy is absorbed and deposited into electrons and ions. The electron energy rises to 4.5 keV during first $4-5$ ps and then remains approximately constant when the interaction enters in a quasi-stationary regime. 

The role of electron collisions is more important in the expanding plasma. Switching the collisions on results in increase of the average electron temperature from 4.5 to 4.7~keV (figure \ref{fig_energy}a), which corresponds to a difference of more than 20\% compared to the total temperature increment $\sim 1$ keV with respect to the initial temperature of 3.5 keV. By contrast, collisions have an opposite effect on the average ion temperature. It increases by 40\% in the collisionless case but only by 30\% in the collisional case. Explanation of the effect of collisions on particle heating is given in the next section.

In the case v) of two ion species shown in figure \ref{fig_energy} with a purple line, the increase of electron temperature is similar to in the case of expanding plasma without collisions. The improved laser absorption in this case is also explained by the SBS suppression, but it is related in this case to a stronger ion acoustic damping on light (hydrogen) ions. The level of SBS reflectivity in this case is 36\%. Consequently, less laser radiation penetrates to the quarter critical density and produces a stronger electron heating.  The effect of hydrogen ions on SRS saturation and hot electron production was reported in several experiments \cite{Fernandez_1996, Kirkwood_1996, Theobald_2017}.

The case vi) of a phase-modulated laser pulse shows a smaller increase of electron and ion temperatures compared to the cases iii) and v). This particle heating is also explained by a partial SBS suppression related to a stochastic nature of laser driver. However, the considered pump correlation time of 0.5 ps is of the same order as the SBS growth time $1/\gamma_{\rm SBS}$ as described in section \ref{sec:theory}. Consequently, the finite bandwidth suppression is less efficient, and the level of SBS reflectivity in this case is maintained on a high level of 81\%. The energy partition in table \ref{tab:table1} shows that laser energy absorption in plasma increases significantly if SBS is suppressed.

\subsection{Laser backscattering}\label{subsec:srs}
\begin{figure}[!ht]
\includegraphics[width=0.32\linewidth]{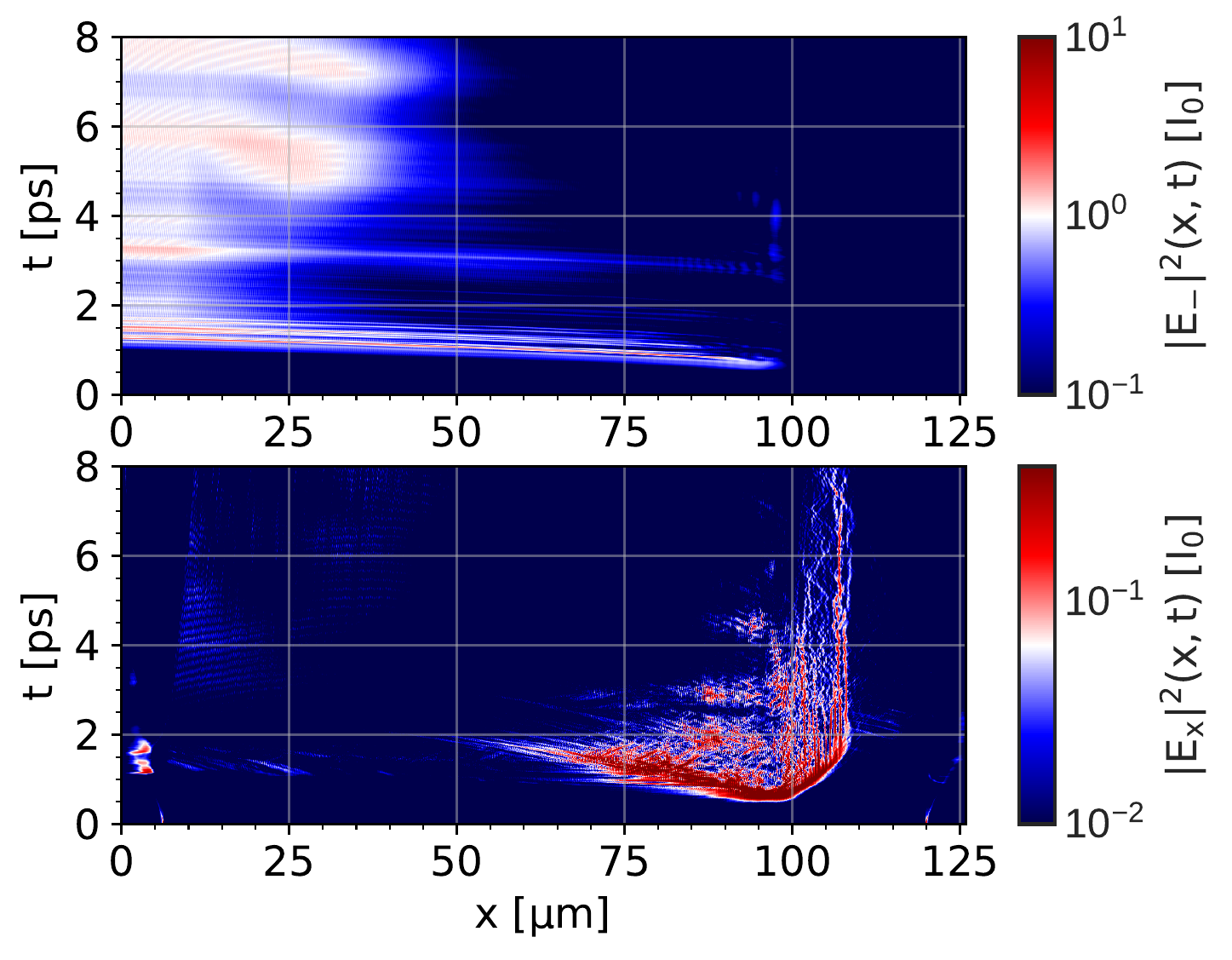}\hspace{1mm}
\includegraphics[width=0.32\linewidth]{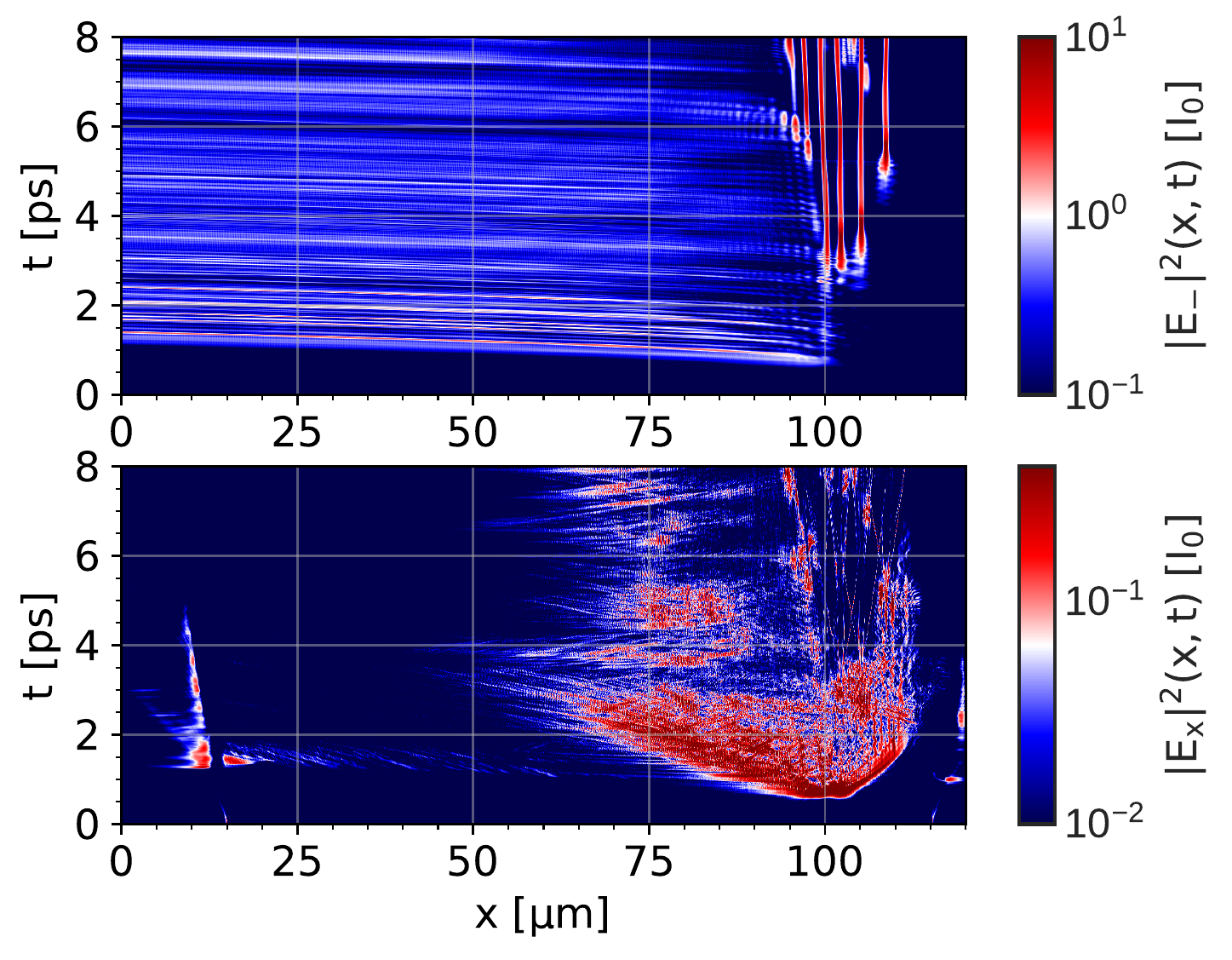}\hspace{1mm}
\includegraphics[width=0.32\linewidth]{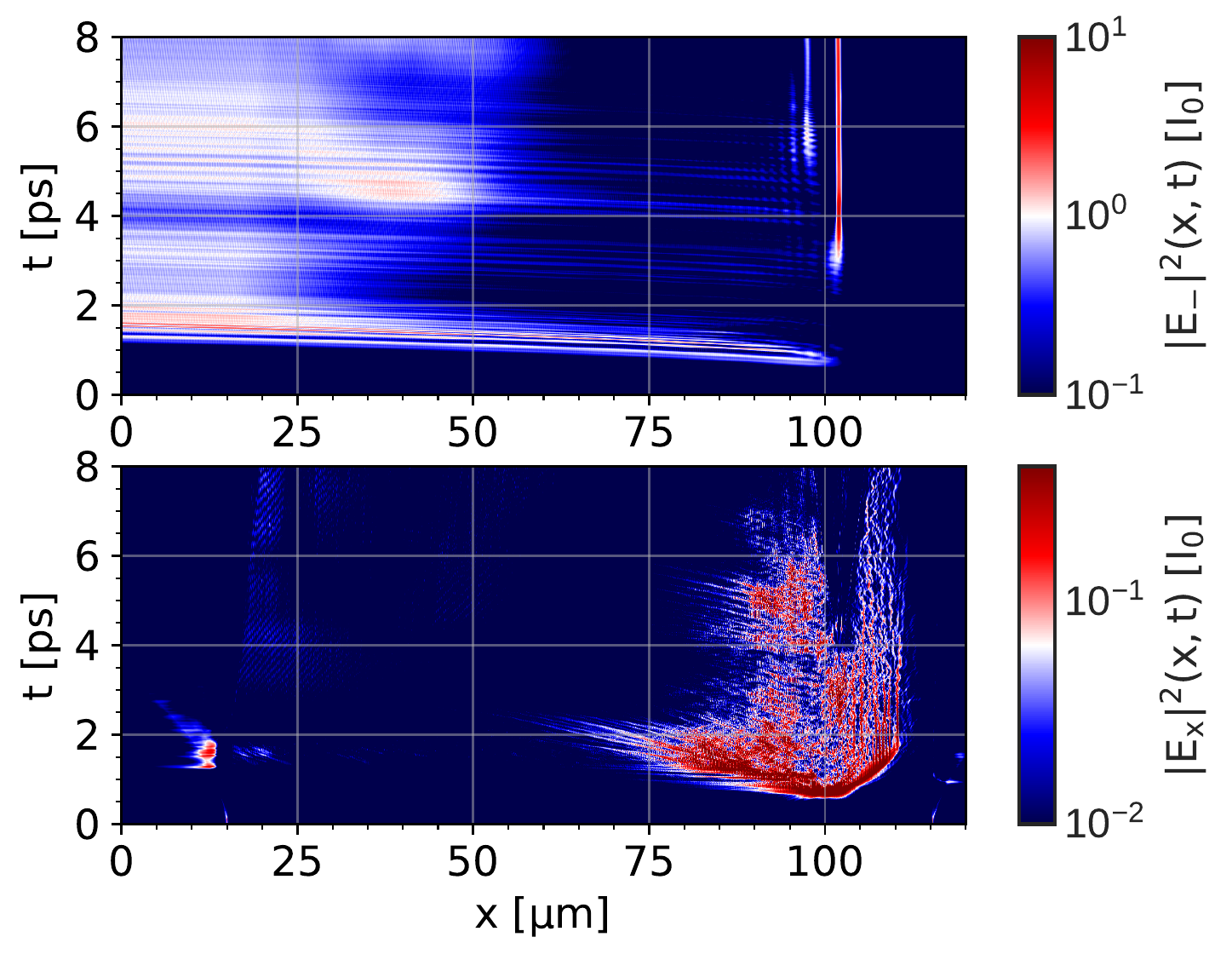}
\caption{Spatial and temporal evolution of the backscattered light (top row) and electrostatic field $E_x$ (bottom row) for the reference case (i, left column), expanding collisionless plasma (iii, central column) and broadband laser pulse (vi, right column).}\label{fig_reflected}
\end{figure}

Both SBS and SRS contribute to the backscattered signal. Their contributions can be distinguished in the spatio-temporal plots in figure \ref{fig_reflected}. Excitation of Raman scattered electromagnetic waves, shown in the top row plots, is correlated with the excitation of large amplitude plasma waves shown in the bottom row. Dynamics of the Raman scattered electromagnetic wave is rather different in the three cases presented in this figure: in the left and right columns corresponding to cases i) and vi), the SBS reflectivity is high, which leads to the inhibition of SRS near the quarter critical density and to suppression of the cavitation process. By contrast, in the central column corresponding to case iii) SBS is suppressed resulting in a stronger SRS and cavitation. 

First short SRS burst comes from the region near the quarter critical density at $t \sim 0.7$ ps. The SRS instability develops very fast, the Langmuir waves are excited in a region near the point $x \simeq 100\,\mu$m, corresponding to $n_e/n_c = 0.23-0.25$, where temporal and spatial gains plotted in figure \ref{fig:gains_SRS} are maximized.  At this time, reflected SRS wave is strong enough to excite secondary SRS near the plasma density of $n_c/16$ at $x\approx 12\,\mu$m manifested by short-lived bursts of plasma waves. This, however, has a very weak effect on the laser reflection and absorption. A steady excitation of SRS is produced later in time $t>4$ ps in the cases iii) and vi), and it is manifested by formation of cavities -- narrow localized packets of electromagnetic field trapped in deep density depressions. There is practically no electrostatic field in the cavities because the electrons are expelled. Formation of plasma cavities near the quarter and one-sixteenths of the critical density due to Raman backscattering was reported by Klimo et al. \cite{Klimo_2010}.

\begin{figure}[!ht]
\includegraphics[width=0.45\linewidth]{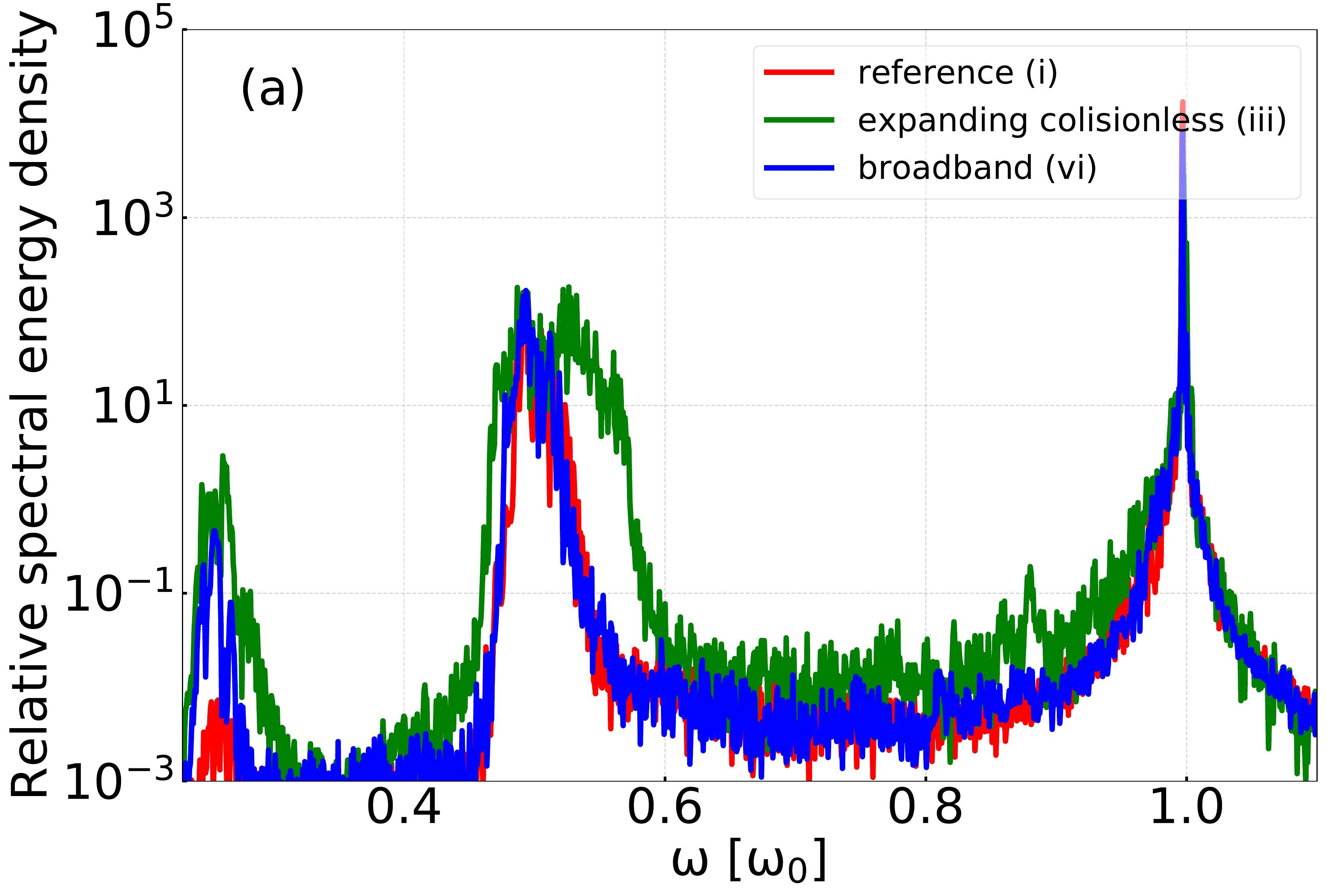}\hspace{1mm}
\includegraphics[width=0.45\linewidth]{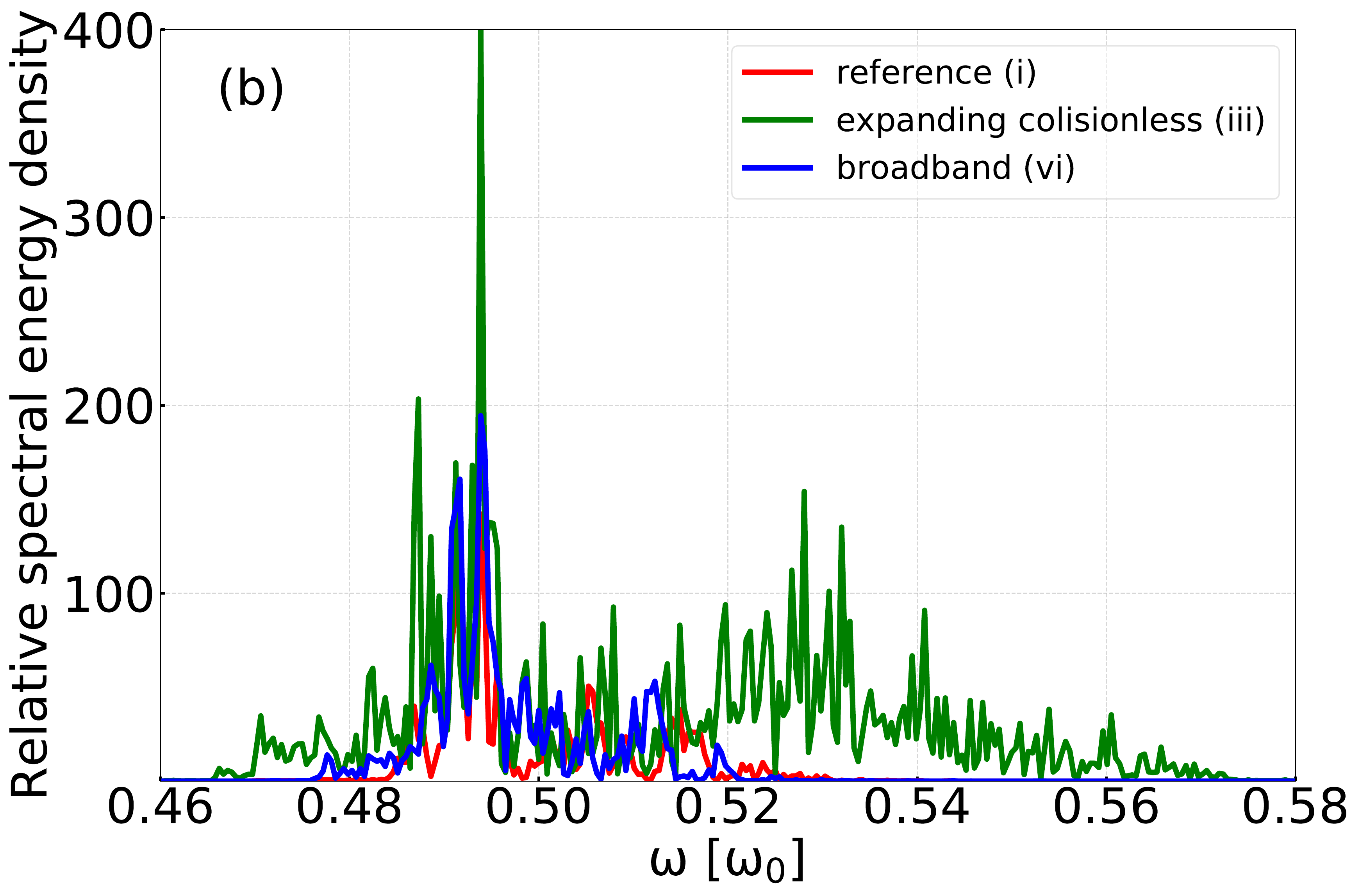}
\caption{Overall view (a) and a detailed structure (b) of the spectrum of the light reflected from the plasma  for the reference case (i, red), expanding collisionless (iii, green) and broadband (vi, blue) cases.}
\label{fig:backscattered_spectrum}
\end{figure}

In the case i) (left column in figure \ref{fig_reflected}), the SRS instability is excited near the quarter critical density only at the transient stage. It produces one or two bursts of scattered light with frequency $\sim 0.5\,\omega_0$ as shown in figure \ref{fig:backscattered_spectrum}a, which could be partially absorbed via collisional damping while propagating through the plasma. Later in time, for $t>4$ ps, the scattered light originates from a plasma with density less than $0.1n_c$, and it is due to SBS. Similar effect is observed in the case vi) of a broadband laser pulse (right column in figure \ref{fig_reflected}). However, here SBS is partially suppressed and the absolute SRS excited at later time, $t>4$ ps, which is manifested in formation of two cavities and stronger backscattering in figure \ref{fig:backscattered_spectrum}. 

\begin{figure}[!ht]
\includegraphics[width=0.45\linewidth]{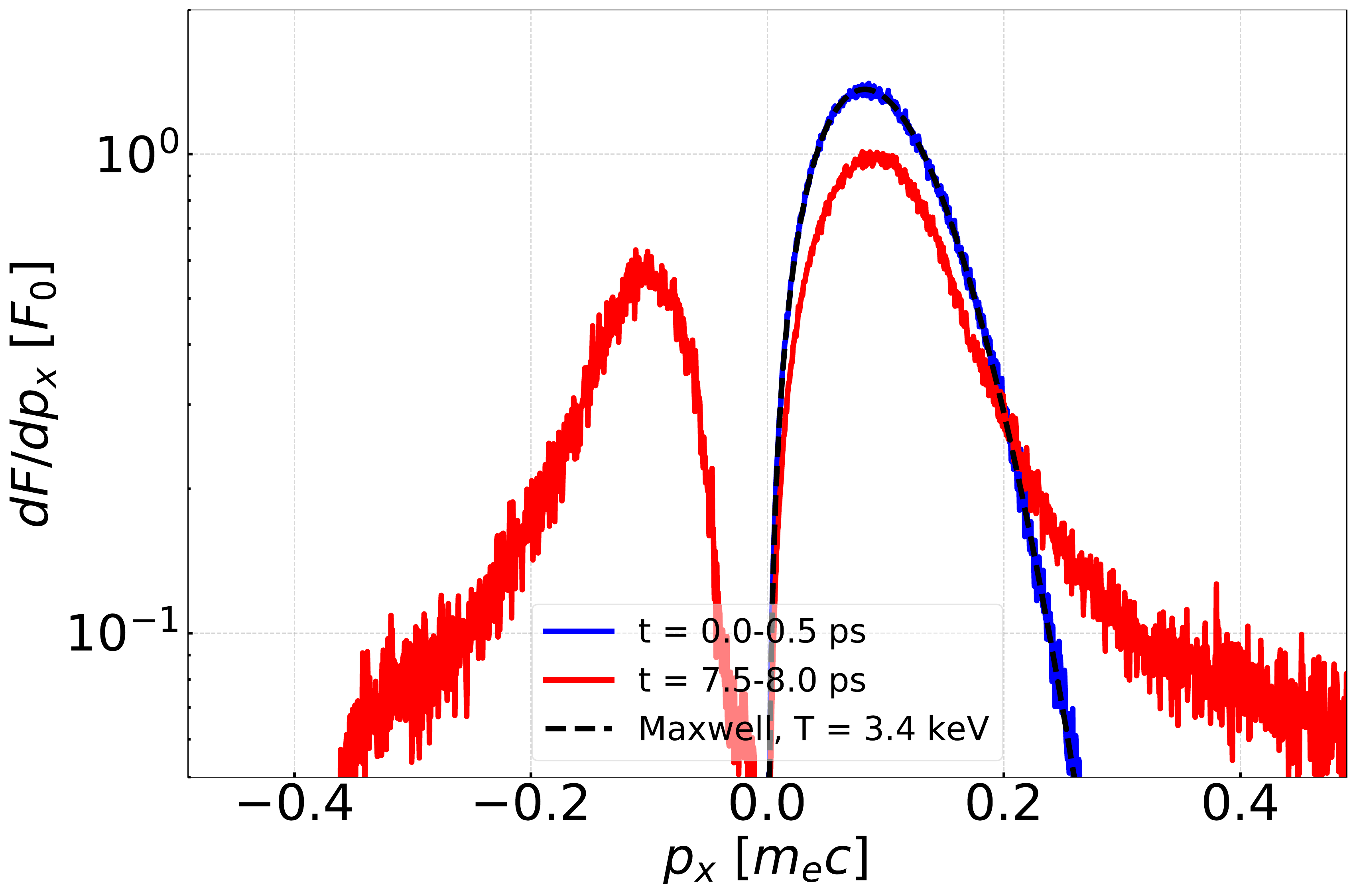}\hspace{1mm}
\caption{Distribution of the electron energy flux near the quarter critical plasma ($n_e \sim 0.25\,n_c$) on the parallel momentum, $dF/dp_x$ averaged over the first (blue) and last 0.5 ps (red) for the case iii) of expanding plasma. Black dashed line is an electron energy flux fitted with Maxwellian function \eqref{distribution_maxwellian} for $p_x > 0$ corresponding to the initial temperature of 3.4 keV. All curves are normalized to the incident laser energy flux $I_0$.}\label{fig:electron_distribution}
\end{figure}

By contrast, SBS is much stronger suppressed in the case iii) of expanding plasma, which is shown in the central column in figure \ref{fig_reflected}. Here, in addition to the absolute SRS, excited near the quarter critical density and produced multiple cavities, a convective SRS is excited in a lower density plasma. The latter is manifested by excitation of Langmuir waves in a plasma with density $\gtrsim 0.15\,n_c$ and a broad spectrum of backscattered waves in figure~\ref{fig:backscattered_spectrum}. 
Strong SRS excitation in the case iii) is accompanied with electron acceleration in the Langmuir waves. The electron energy flux shown in figure \ref{fig:electron_distribution} (red) could be fitted with a Maxwellian two-temperature distribution:
\begin{equation}\label{distribution_maxwellian}
    \frac{dF}{dp_x} = \sum_{j = {\rm bulk,\, hot}} \frac{n_j\,v_x }{\sqrt{2\pi m_e T_{e\,j}}}\Bigl(\frac{p_x^2}{2m_e}+T_{e\,j}\Bigr) \exp\Bigl(-\frac{p_x^2}{2m_eT_{e\,j}}\Bigr),
\end{equation}
with temperatures of bulk $T_{e\,\rm bulk} \simeq 3.5$ keV and hot $T_{e\,\rm hot} \simeq 35$ keV components, respectively. The ratio of hot to bulk relative concentrations $n_j$ is about 10.7\%. As the bulk electrons are reflected from the vacuum zone to the right of plasma edge at $x\simeq 120$, the electron distribution function is approximately symmetric and only part corresponding to $p_x>0.3\,m_ec$ is absorbed at the right boundary.

The interpretation of respective roles of SRS and SBS in reflection of the incident laser light is corroborated by analysis of the spectrum of reflected light in figure \ref{fig:backscattered_spectrum}b. The peak near the laser frequency dominates the spectrum in simulation cases i), vi). It is downshifted by $\Delta\omega_a \simeq 0.0028\,\omega_0$, which is consistent with the frequency of a driven ion wave $\omega_{ia} = 2k_0 c_s$. Spectral downshift in the two species case v) $\Delta\omega_a \simeq 0.0032\,\omega_0$ is consistent with excitation of the fast wave, as discussed in section \ref{sec:sbs} 

A particular feature of SRS spectrum shown in figure \ref{fig:backscattered_spectrum} is that it extends to the frequencies smaller than $\omega_0/2$. It contains several spectral components but the strongest one is downshifted by $\Delta\omega_s=\omega_s-\omega_0/2 \approx -0.01\omega_0$. This shift was predicted theoretically in Ref. \cite{Afeyan_1985}. It is explained by thermal electron motion, which leads to effective increase of the plasma wave frequency and to a correspondent decrease of the scattered wave frequency. According to equation (56) in Ref. \citep{Afeyan_1985}, expression for the frequency shift of the scattered wave reads: 
\begin{equation}\label{eq5}
\frac{\Delta\omega_s}{\omega_0} = \frac{\omega_s-\omega_0/2}{\omega_0} \simeq -\frac{9T_e}{8m_ec^2}+\frac{1}{4k_0L_n(v_{os}/c)^{1/2}}.
\end{equation}
The first term in the right hand side is negative and it is on the order of 0.01 under our conditions, which agrees very well with the observed shift. The second term is much smaller, it is on the order of 0.001. A similar negative shift was observed in several other publications \citep{Klimo_2014, Gu_2019}, but its origin was not explained. 

Formation of cavities is related to the fact that the group velocity of the scattered electromagnetic wave near the quarter critical density is close to zero and thereby it stays a relatively long time in the resonance with the pump wave \cite{Liu_1974, Afeyan_1985}. Consequently, a strong ponderomotive force produced by the daughter electromagnetic and plasma waves induces a density depression where they are self-trapped. The plasma wave disappears when electrons are expelled from the cavity, while electromagnetic waves are weakly damped and persist for a long time. Ions are accelerated under the ponderomotive pressure of trapped light as discussed in section \ref{subsec:cavity}. The cavity formation is manifested by a strong reduction of backscattering level to $\sim25$\%.

The cavity formation is terminated when the ponderomotive pressure of trapped waves is equalized  with the thermal pressure of the ambient plasma. So, the energy stored in the cavity can be estimated as $0.25 n_cT_e \Delta x_w$, which corresponds to about 1 kJ/cm$^2$ for our conditions. The cavity life time is defined by the damping of the trapped mode, which is about 1 ps according to equation  \eqref{cavity_absorption}. Therefore, cavitation and subsequent collisionless absorption of trapped electromagnetic waves could be an important mechanism of laser energy absorption as one can see in table \ref{tab:table1}.

Collisional absorption of the scattered electromagnetic wave makes also a notable contribution to the absorption process. This can be seen in figure \ref{fig_energy} when comparing the cases with and without collisions. In the reference case, there is no difference between the runs i) and ii): since the SRS is suppressed by a strong SBS in a low density plasma. By contrast, there is a notable electron and ion heating in the case of expanding plasma: in the run iii) without  collisions, both electrons and ions are heated due to SRS excitation and energy absorption in cavities. In the run iv) with collisions, there is less ion heating and more electron heating due to the collisional absorption of the SRS scattered wave.

\subsection{Ion heating and cavitation}\label{subsec:cavity}
Excitation of parametric instabilities leads also to the ion heating. It is related to two processes: damping of ion acoustic waves excited by SBS in a low density plasma, $n_e/n_c\lesssim 0.2$, and expansion of cavities created at nonlinear stage of SRS evolution near the quarter critical density. A comparison of the ion temperature evolution for different cases in figure \ref{fig_energy}b shows that cavitation makes a strong contribution with ion temperature increase by more than 30\% during the simulation time.

\begin{figure}[!ht]
\includegraphics[width=0.45\linewidth]{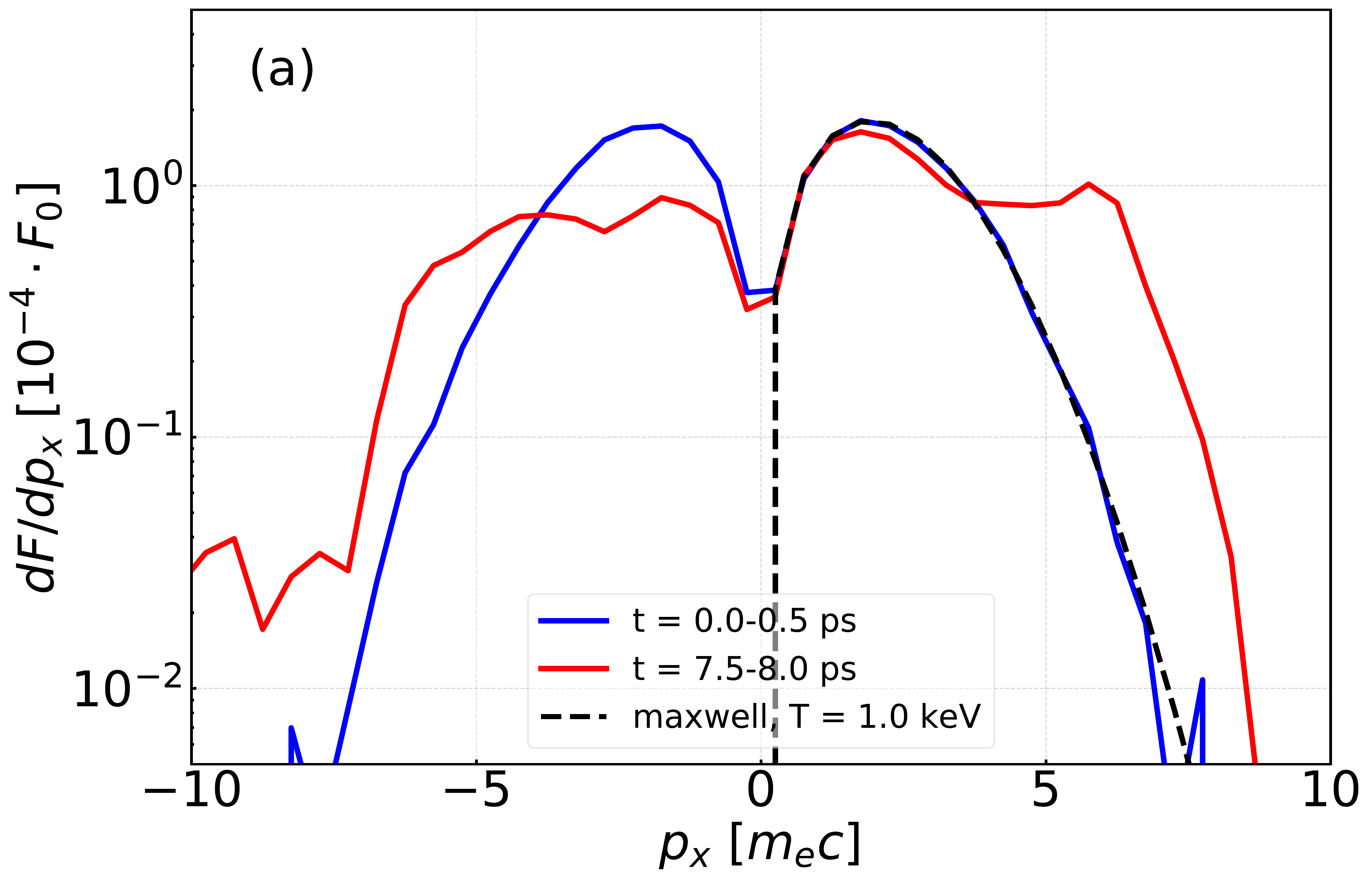} \hspace{3mm}
\includegraphics[width=0.45\linewidth]{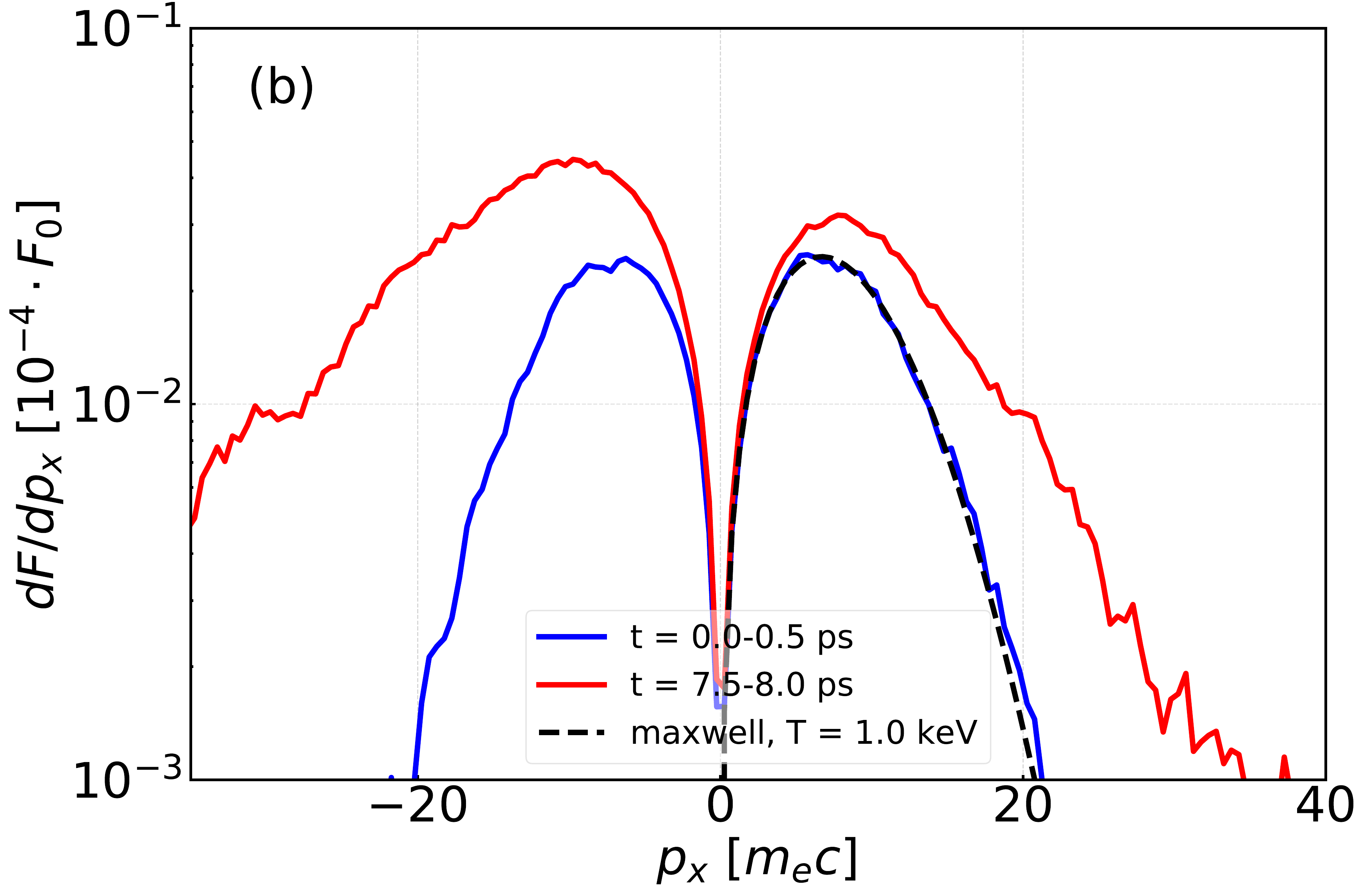}
\caption{Distribution of the ion energy flux near the quarter critical plasma on the parallel momentum, $dF/dp_x$ averaged over the first (blue) and last 0.5 ps (red) for the case v) with two ion species: (a) hydrogen and (b) carbon. Black dashed line is an ion energy flux fitted with Maxwellian function \eqref{distribution_maxwellian} for $p_x > 0$ corresponding to the initial temperature of 1 keV. All curves are normalized to the incident laser energy flux $I_0$. Total ion energy flux through the quarter critical plasma is about $1.6\times 10^{-3}I_0$ for hydrogen and $\sim 10^{-4}I_0$ for carbon.
} \label{fig_distribution1}
\end{figure}

\begin{figure}[!ht]
\includegraphics[width=0.45\linewidth]{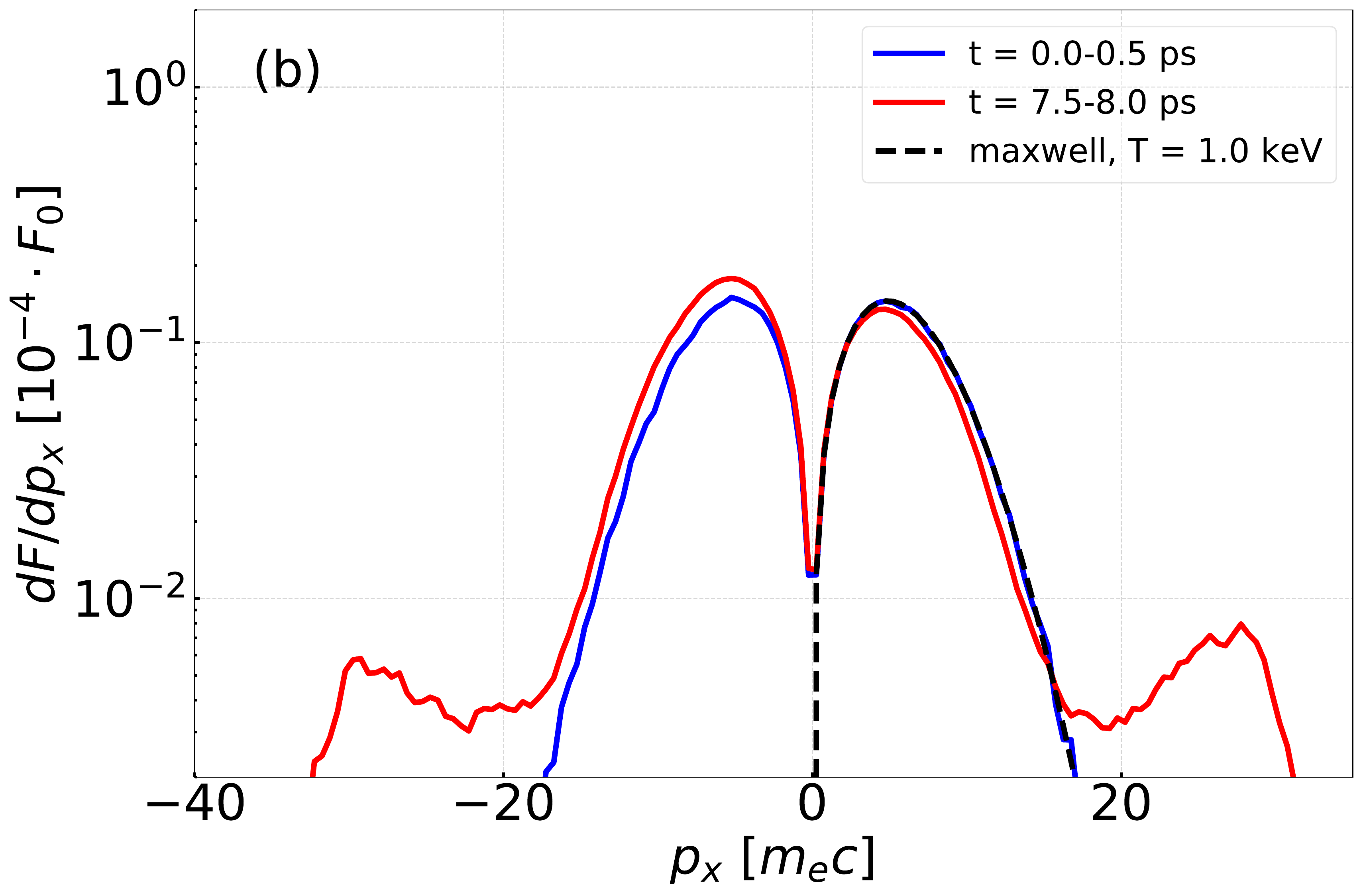}\hspace{3mm}
\includegraphics[width=0.45\linewidth]{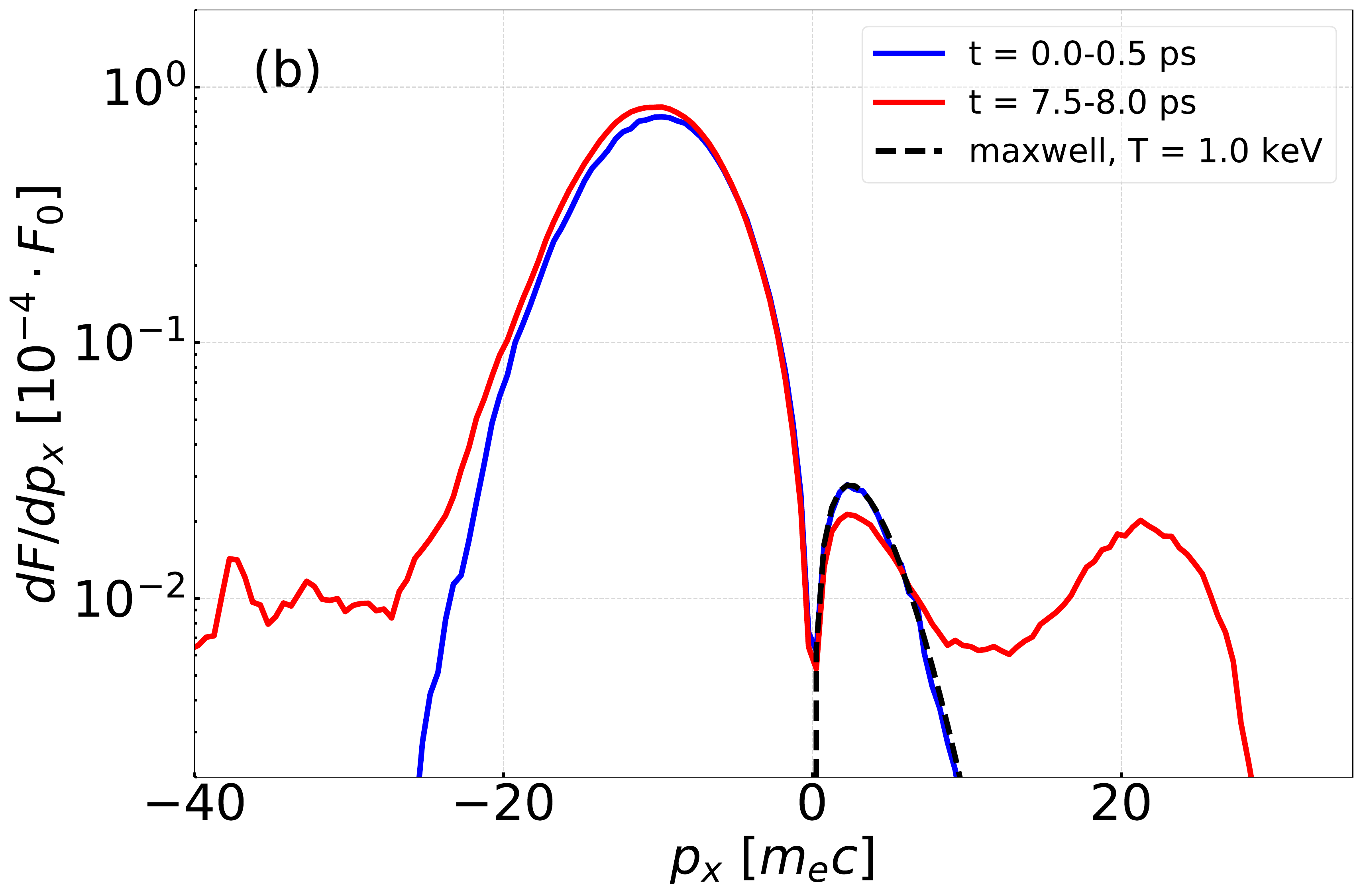}
\caption{Distribution of the ion energy flux entering the quarter critical plasma on the parallel momentum, $dF/dp_x$ averaged over the first (blue) and last 0.5 ps (red) for broadband laser pulse vi) (a) and for expanding plasma iii) (b). Black dashed line is an ion energy flux fitted with Maxwellian function \eqref{distribution_maxwellian} for $p_x > 0$ corresponding to the initial temperature of 1 keV. All curves are normalized to the incident laser energy flux $I_0$.
} \label{fig_distribution2}
\end{figure}

A more detailed analysis of the ion dynamics is presented in figures \ref{fig_distribution1} and \ref{fig_distribution2} showing distribution of the ion energy flux on the momentum $p_x$ near the quarter critical density. Figure \ref{fig_distribution1} shows the case v) of two ion species. According to \cite{Bychenkov_1995a, Williams_1995}, for the temperature ratio, $T_i/T_e\simeq 0.3$, the fast mode has a smaller damping increment, ${\rm Im}\,\omega_{ia}/{\rm Re}\,\omega_{ia} \simeq 0.24$, and it is excited in the SBS process. The phase velocity of  this mode $\simeq 0.76\,\mu$m/ps is approximately two times the hydrogen thermal velocity and it is approximately eight times the carbon thermal velocity.  Consequently, a comparison of the left and right panels in figure \ref{fig_distribution1} shows that hydrogen ions are gaining much more energy than carbon ions from the ion acoustic waves.

SBS develops in the case near the quarter critical density. Indeed, the distribution function of both ion species is modified essentially for positive velocities and in the range larger than the ion acoustic velocity. The cutoff of the proton distribution at $p_x\simeq 10\,m_ec$ corresponds to the proton energy of 12 keV and velocity approximately three times the phase velocity of the slow mode. Velocity of carbon ions is about 5 times smaller than the hydrogen velocity and they are carrying an order of magnitude smaller energy flux.

Figure \ref{fig_distribution2} shows the ion energy flux for two other simulations with a single ion species with an expanding plasma iii) and a broadband laser pulse vi). SBS is suppressed in these two runs and modification of the ion distribution function is completely different. The ions are accelerated symmetrically in the positive and negative directions and their velocities are 1.5 times larger than the ion acoustic velocity $c_s \simeq 0.5\,\mu$m/ps, which corresponds to the ion momentum $\sim 20\,m_ec$. This symmetric ion acceleration is explained by their expansion from the cavities produced at the nonlinear stage of SRS evolution. This interpretation is confirmed by the reduction of backward reflectivity and electron heating observed in these runs.

\section{Conclusion}\label{conclusion}
We discussed and compared several methods of controlling laser absorption in a hot, weakly collisional plasma via SRS-SBS competition. The reference case of a plasma with zero fluid velocity and a single average ion species is characterized by a relatively week laser absorption and strong  SBS backscattering. This situation corresponds qualitatively to experiments with single ion species plasmas \cite{Theobald_2017} showing low level SRS saturation and small number of hot electrons.

Such a situation is not typical for experiments with solid density targets where underdense plasma is expanding with a supersonic velocity. It is important to account for this factor in numerical simulations of laser plasma interaction. Divergence of the plasma flow velocity leads to a strong reduction of the SBS spatial gain and makes a dramatic effect on the SBS-SRS competition. The SBS reflectivity decreases by a factor of three, thus allowing laser to penetrate to the quarter critical density region and excite absolute SRS instability. The leads to a significant nonlinear laser absorption through the cavitation and collisional absorption of the scattered electromagnetic wave, which could be as big as 30\%. Hot electron temperature observed in our simulations does not exceed 50 keV, which is compatible with the shock ignition scenario of inertial fusion.

Two other methods of suppression of a strong SBS amplification are the use of a multi-species plasma and/or a broadband laser pulse. By adding hydrogen ions with the thermal velocity close to the phase velocity of ion acoustic waves one can dramatically reduce the SBS gain and favor SRS excitation near the quarter critical density. The SBS-SRS competition in this case is manifested in a very different modification of the ion distribution function. Instead of ion acceleration in the forward direction in large amplitude ion acoustic waves, SRS and cavitation manifest themselves in a symmetric ion acceleration in the plasma cavities. 

A phase-modulated laser pulse can also suppress SBS and favor SRS, but the considered case with a correlation time $\tau_c \sim 0.5$ ps is too large thus leading to a relatively small reduction of reflectivity. It is desirable to increase the laser bandwidth at least $2-3$ times for a more efficient SBS suppression. 

To summarize, we revised a set of flexible methods for controlling the SRS-SBS competition in inertial fusion plasmas. Suppression of SBS reflectivity favors absolute and convective SRS excitation, electron acceleration and cavitation. The latter effect could make an important contribution to the laser  absorption by trapping the backscattered light. 

\section*{Acknowledgments}
Fruitful discussions with S. Weber are gratefully acknowledged. This research was partially supported by the Project LQ1606 with the financial support of the Ministry of Education, Youth and Sports as part of targeted support from the Czech National Programme of Sustainability II.

\bibliographystyle{apsrev}
\bibliography{paper1_SS}

\end{document}